\documentclass[epsf,11pt]{article}
\usepackage{aaspp4,lscape,psfig}
\tighten

\def\gtsima{$\, \buildrel > \over \sim \,$}
\def\ltsima{$\, \buildrel < \over \sim \,$}
\def\simgt{\lower.5ex\hbox{\gtsima}}
\def\simlt{\lower.5ex\hbox{\ltsima}}

\begin{document}
 
\title{Mass-Losing Semiregular Variable Stars in Baade's Windows}

\medskip
\medskip


\author{
      C.~Alard\altaffilmark{1,12}, 
      J.A.D.L.~Blommaert\altaffilmark{2}, 
      C.~Cesarsky\altaffilmark{3},
      N.~Epchtein\altaffilmark{4},
      M.~Felli\altaffilmark{5},
      P.~Fouque\altaffilmark{6},\\
      S.~Ganesh\altaffilmark{7},
      R.~Genzel\altaffilmark{8},
      G.~Gilmore\altaffilmark{9}, 
      I.S.~Glass\altaffilmark{10},
      H.~Habing\altaffilmark{11},
      A.~Omont\altaffilmark{12},
      M.~Perault\altaffilmark{13},\\
      S.~Price\altaffilmark{14},
      A.~Robin\altaffilmark{15},
      M.~Schultheis\altaffilmark{12},
      G.~Simon\altaffilmark{1},
      J.Th.~van Loon\altaffilmark{9},\\
      {\it (The ISOGAL Collaboration)}
}

\author{
      C.~Alcock\altaffilmark{16,30}, 
      R.A.~Allsman\altaffilmark{17},
      D.R.~Alves\altaffilmark{18},
      T.S.~Axelrod\altaffilmark{19},
      A.C.~Becker\altaffilmark{20},\\
      D.P.~Bennett\altaffilmark{21,30},
      K.H.~Cook\altaffilmark{16,30},
      A.J.~Drake\altaffilmark{16},
      K.C.~Freeman\altaffilmark{19},
      M.~Geha\altaffilmark{22},\\
      K.~Griest\altaffilmark{23,30},
      M.J.~Lehner\altaffilmark{24},
      S.L.~Marshall\altaffilmark{16},
      D.~Minniti\altaffilmark{25}, 
      C.~Nelson\altaffilmark{26},
      B.A.~Peterson\altaffilmark{19},\\
      P.~Popowski\altaffilmark{16},
      M.R.~Pratt\altaffilmark{27},
      P.J.~Quinn\altaffilmark{3},
      W.~Sutherland\altaffilmark{28,30},
      A.B.~Tomaney\altaffilmark{26},
      T.~Vandehei\altaffilmark{23},
      D.L.~Welch\altaffilmark{29}\\
      {\it (The MACHO Collaboration)}
}


\vskip3cm


{\footnotesize
%
%
\altaffiltext{1}{DASGAL, UMR CNRS N$_o$ 335, Observatoire de Paris,          
        21 Avenue de l'Observatoire, Paris F75014, France }                   
\altaffiltext{2}{ISO Data Centre, Astrophysic Div., Space Science            
        Dept.~of ESA, Apartado 50727, Madrid E28080, Spain }                  
\altaffiltext{3}{ESO, Karl-Schwarzschild-Str 2, D-85740, Germany}            
\altaffiltext{4}{Observatory de la Cote d'Azur, Departement Fresnel,         
BP 4229, F 06304 Nice Cedex 4, France}                                        
\altaffiltext{5}{Observatory of Arcetri, Largo E. Fermi 5, 50125, Florence,  
        Italy}                                                                
\altaffiltext{6}{ESO, Casilla 19001, Santiago, Chile}                        
\altaffiltext{7}{Physical Research Laboratory, Astronomy and Astrophysics    
        Division, Navarangpura, Ahmedabad 380 009, India }                    
\altaffiltext{8}{MPE, Karl-Schwarzschild-Str 1, Postfach 1603, D-85740        
        Garching, Germany}                                                    
\altaffiltext{9}{Institute of Astronomy, The Observatories,                  
        Madingley Rd., Cambridge CB3 0HA, United Kingdom }                    
\altaffiltext{10}{South African Astronomical Observatory,                     
        P.O. Box 9, Observatory 7935, South Africa}                           
\altaffiltext{11}{Sterrewacht Leiden, P.O. Box 9513, 2300 RA Leiden, 
        The Netherlands}                                                       
\altaffiltext{12}{Intitut d'Astrophysique de Paris, CNRS,                    
        98bis Blvd Arago, Paris F75014, France}                               
\altaffiltext{13}{Laboratoire de Radioastronomie Millim\'{e}trique, ENS \&    
        CNRS, 24 Rue Lhomond, F-75005 Paris, France}                          
\altaffiltext{14}{AFRL, Hanscom AFB, MA, 01731 USA}                
\altaffiltext{15}{Observatoire de Besan\c{c}on, BP 1615, F-25010              
        Besan\c{c}on Cedex, France}                                           
}                                                                             
%
%

\altaffiltext{16}{Lawrence Livermore National Laboratory, Livermore, CA 
        94550, USA}
\altaffiltext{17}{Supercomputing Facility, Australian National University,
        Canberra, ACT 0200, Australia }
\altaffiltext{18}{Space Telescope Science Institute, 3700 San Martin Dr.,
        Baltimore, MD 21218, USA}
\altaffiltext{19}{Mt.~Stromlo and Siding Spring Observatories, ANU,
        Weston Creek, ACT 2611, Australia}
\altaffiltext{20}{Departments of Astronomy \& Physics,
        University of Washington, Seattle, WA 98195, USA}
\altaffiltext{21}{Physics Department, University of Notre Dame, Notre 
        Dame, IN 46556, USA}
\altaffiltext{22}{Department of Astronomy, University of California,
        Santa Cruz, CA 95064, USA}
\altaffiltext{23}{Department of Physics, University of California,
        San Diego, La Jolla, CA 92093, USA}
\altaffiltext{24}{Department of Physics, University of Sheffield,
        Sheffield S3 7RH, United Kingdom }
\altaffiltext{25}{Departmento de Astronomia, P. Universidad Catolica, 
        Casilla 104, Santiago 22, Chile}
\altaffiltext{26}{Department of Physics, University of California,
        Berkeley, CA 94720, USA }
\altaffiltext{27}{Center for Space Research, MIT, Cambridge, MA 02139, USA}
\altaffiltext{28}{Department of Physics, University of Oxford,
        Oxford OX1 3RH, U.K. }
\altaffiltext{29}{Departments of Physics \& Astronomy, 
   McMaster University, Hamilton, Ontario, Canada L8S 4M1. }
\altaffiltext{30}{Center for Particle Astrophysics,
        University of California, Berkeley, CA 94720, USA}


\clearpage

\begin{abstract}

By cross-correlating the results of two recent large-scale surveys, 
the general properties of a well defined sample of semi-regular 
variable stars have been determined. ISOGAL mid-infrared photometry 
(7 and 15 $\mu$m) and MACHO $V$ and $R$ lightcurves are 
assembled for approximately 300 stars in the Baade's Windows of 
low extinction towards the Galactic bulge.  These stars are mainly 
giants of late M spectral type, 
evolving along the asymptotic giant branch (AGB).
They are found to possess a wide and continuous distribution 
of pulsation periods and to obey an approximate log~period -- bolometric
magnitude (log\,$P$ -- $M_{\rm bol}$) relation or set of such relations. 

Approximate mass-loss rates $\dot{M}$ in the range of $\sim$1 $\times$ 
10$^{-8}$ to 5 $\times$ $10^{-7}$ $M_{\odot}$~year$^{-1}$ are derived 
from ISOGAL mid-infrared photometry and models of stellar spectra 
adjusted for the presence of optically-thin circumstellar silicate dust.  
Mass-loss rates depend on luminosity and pulsation period. 
Some stars lose mass as rapidly as short-period Miras but do not 
show Mira-like amplitudes. A period of 70 days or longer is a necessary 
but not a sufficient condition for mass loss to occur.

For AGB stars in the mass-loss ranges that we observe, the functional 
dependence of mass-loss rate on temperature and luminosity can be 
expressed as $\dot{M} \propto T^{\alpha}L^{\beta}$,
where $\alpha=-8.80^{+0.96}_{-0.24}$ and $\beta=+1.74^{+0.16}_{-0.24}$,
in agreement with recent theoretical predictions. 

If we include our mass-loss rates with a sample of extreme mass-losing
AGB stars in the Large Magellanic Cloud, 
and ignore $T$ as a variable, we get the general result for AGB stars that
\[                                                                         
\dot{M} \propto L^{2.7},                                                   
\]                                                                         
valid for AGB stars with 10$^{-8}$  $<$ $\dot{M}$ $<$ 10$^{-4}$ $M_{\odot}$ 
yr$^{-1}$. 
                                                                           
\end{abstract}

\clearpage

\section{Introduction}

One of the most complex and least understood phases of stellar
evolution is the asymptotic giant branch (AGB; Iben \& Renzini 1983).
AGB evolution is regulated by very high rates of mass loss 
(Bowen \& Willson 1991; Vassiliadis \& Wood 1993).

The current theory of mass loss from red giants
invokes a combination of two physical processes:
stellar pulsation and radiation pressure on dust grains
(see reviews by Morris 1987; Gail, Kuntz \& Ulmschneider 1990; Lafon 
\& Berruyer 1991; Habing 1996).
The association of mass loss with pulsation is due to the 
fact that mass-losing red giants are often variable stars, such as 
Miras, and the mass-loss rates of Miras
are observed to be correlated with their pulsation periods
(de~Gioia-Eastwood et al.~1981; Whitelock et al.~1994, 1995).
Stellar pulsation is theoretically linked to mass loss through
the propagation of periodic shocks (Wood 1979; Willson \& Hill 1979).
The shocks are nearly isothermal, and thus responsible by themselves for very
low mass-loss rates (Bowen 1988).  However, they also extend the
outer-atmospheres of red giants, increasing the gas density at 
the dust condensation radius.  Dust therefore forms efficiently, and 
radiation pressure accelerates the dust grains (Hoyle \& Wickramasinghe 1962).
The accelerated dust grains are momentum-coupled to the gas, which
drives mass loss at high rates (Gilman 1972; Kwok 1975).
The standard theory of mass loss has evolved to include sophisticated
treatments of time-dependent hydrodynamics, grain condensation, and
radiative transfer (e.g.~Fleischer, Gauger \& Sedlmayr 1992;
Arndt, Fleischer \& Sedlmayr 1997).

The standard theory of mass loss in red giants may not 
be appropriate for all AGB stars.  While the majority are 
variable (e.g.~Alcock et al.~2000), most do not show the 
large-amplitude, long-period pulsations characteristic of Miras.  
Instead, most AGB variable stars are classified as semiregulars (SRs).
The classical requirement for a Mira is that it shows an optical pulsation
amplitude $\Delta$V~$>$~2.5~mag, while the semiregulars are defined
as having pulsation amplitudes smaller than this. Stars with visual
amplitudes around the dividing level occur relatively infrequently
(Payne-Gaposchkin, 1951). 
The optical pulsation amplitudes of current model AGB stars with periodic
shocks as prescribed by the theory of mass loss are typical of the 
pulsation amplitudes of Miras, but not of semiregulars.
Another characteristic feature of Miras is periodic Balmer-line emission, 
believed to arise from shocks (Willson 1976).  The theoretical velocity 
changes across the periodic shocks are in agreement with those inferred 
from the Balmer-line emission of Miras (Fleischer, Gauger \& Sedlmayr 
1992).  Although data are scarce, many SRs do not exhibit periodic 
Balmer-line emission.  In addition, near-infrared spectra of SRs do not 
show line-doubling as Miras often do, a characteristic of shocks (Hinkle,
Lebzelter, \& Scharlach 1997). Thus the pulsations induced in the standard 
theory of mass loss are probably too strong to be appropriate for 
semiregulars.

Up till now, detailed knowledge of the properties of the SR 
variables has been limited to relatively small numbers of bright objects, 
mainly situated in the solar neighborhood at unknown distances
(e.g.~Jura \& Kleinmann 1992; Kahane \& Jura 1994;
~Kerschbaum, Olofsson \& Hron 1996; Mennessier et al.~2000). However, 
the picture is undergoing rapid change thanks to the gravitational lensing 
experiments like MACHO and OGLE, and the astrometric satellite Hipparcos,
which have obtained frequent photometric measurements of large samples 
of SRs at known distances, and over long periods of time.
These relatively new databases cover well-defined samples and 
can reveal variations with amplitudes as small as a few hundredths of a 
magnitude, well beyond the capabilities of earlier photographic work. 
We mention Alves et al.\ (1998), Minniti et al.\ (1998), Wood et al.\ (1999), 
and Glass et al.\ (2000) who have discussed MACHO observations of SRs in 
the LMC and Galactic Bulge.  Koen \& Laney (2000), 
Bedding \& Zijlstra (1998), and others have discussed SRs observed 
by Hipparcos in the solar neighborhood.

At the same time, the ISOCAM camera of the 
Infrared Space Observatory (ISO)\footnote{ISO is a 
European Space Agency (ESA) project with instruments funded by member 
states (especially the PI countries: France, Germany, the Netherlands, 
and the United Kingdom) and with the participation of ISAS and NASA.} 
pointed, observatory-style satellite has enabled mid-infrared photometric 
surveys to be carried out with much greater sensitivity and spatial 
resolution than, for example, was possible with IRAS, which suffered 
severely from crowding of sources near the Galactic plane. 
In particular, in the Baade's Windows of low extinction in the inner 
part of the Galactic bulge, the ISOGAL Collaboration
has detected 1193 stars in the ISOCAM 7$\mu$m or $15\mu$m bands in two 
fields of 15 $\times$ 15 arcmin$^2$ (Glass et al.~1999).  As a
result, it is now evident that there is a continuous sequence of increasing 
mass-loss from mid- to late-type M-giant stars on the AGB,
ending with the Miras and other long-period, large-amplitude 
variables.

In order to advance our theories of AGB stellar evolution, 
and in particular, advance our understanding of mass loss, we 
have undertaken a new study of AGB stars in the Galactic bulge.    
We have combined two types of observations ideally suited to 
investigate issues of AGB star mass loss: optical-band lightcurves 
from the MACHO Project and mid-infrared photometry from the ISOGAL 
Collaboration. Our dataset is the first large sample 
of mass-losing AGB variable stars whose pulsation and mass-loss rates
are well-characterized, and whose distances, and thus energetics,
are also known. 

\section{Data}

\subsection{ISOGAL}

The ISOGAL Survey\footnote{This is paper no.~10 in a refereed journal
based on data from the ISOGAL Survey.}
is a multi-wavelength infrared survey of the inner 
Galaxy at high resolution (Omont et al.~2000). 
It made use of the ISOCAM camera (Cesarsky et 
al.~1996) on the ISO satellite to survey numerous 
sample fields in visually-obscured regions along the Galactic plane 
and towards the center of the Galaxy in order to study topics such as 
Galactic structure, the red giant population, interstellar extinction 
and other matters.  

\subsubsection{ISOGAL observations}

The survey comprises mainly exposures in the LW2 (5.5--8$\mu$m) and 
LW3 (12--18$\mu$m) broad-band filters of ISOCAM. Each pixel subtended 6 
$\times$ 6 arcsec$^2$ on the sky. The detector arrays had 32 $\times$ 32
pixels. Large areas could be imaged by combining individual images obtained
during raster scans. The data that we describe were obtained from two
areas located in the Baade's Windows, each covering 15 $\times$ 
15 arcmin$^2$ in $\ell,b$. Exposures were made on two occasions, about 
a year apart (see Glass et al.\ 1999). On the first occasion, observations 
were obtained only with the LW2 filter; on the second, both filters were 
used. The centres of
the fields were at $\ell = + 1.03^{\circ}$, $b= -3.83^{\circ}$, which
includes the globular cluster NGC\,6522, and at $\ell = + 1.37^{\circ}$, 
$b= -2.63^{\circ}$, known as the SgrI field. The Baade's Window fields 
were included because, although they are near the galactic centre, they 
are sufficiently unobscured (with visual extinction $A_V$ $\sim$ 1.5 -- 1.8 
mag) to be observable 
at visual wavelengths and have been the subject of numerous previous 
investigations. Thus they can be regarded as fiducial fields for the 
analysis of more heavily obscured areas.  

\subsubsection{Photometric properties}

ISOGAL has identified 1193 sources from these fields at either or both of 
the 7- and 15-micron passbands.  In the fields Sgr\,I and NGC\,6522, there are
696 and 497 sources, respectively. Of these, 182 and 287 were detected 
at both wavelengths. The methods used for reducing the photometry, taking
account of various difficulties produced by the responses of the detectors
and the crowded nature of the fields, have been described by Glass et al.\ 
(1999). The flux calibrations of the 7\,$\mu$m and 15\,$\mu$m bands were 
set for a spectrum with $F_{\lambda} \propto \lambda^{-1}$ and should be 
correct at wavelengths 6.7 and 14.3\,$\mu$m respectively. Conversion from
magnitudes follows the relations
\begin{equation}
[7]= 12.38-2.5\;{\rm log}F_{LW2}(\rm mJy)
\end{equation}
and
\begin{equation}
[15]= 10.79-2.5\;{\rm log}F_{LW3}(\rm mJy),
\end{equation}
where the zero points have been set to obtain zero mag for a Vega model flux
at the isophotal wavelengths given above.

Only sources with fluxes greater than 5\,mJy in either or both filters were
accepted for the final catalogue. These limits correspond to [7] = 10.64 and
[15] = 8.99. The rms dispersion of the
ISOGAL photometry is 0.14--0.2\,mag, except at the
faint end, where it rises to $\sim$ 0.4\,mag (see Glass et al.\ 1999; Ganesh
et al.\, in preparation). With 35--45 pixels per source, observations in
these fields are seen to be close to the confusion limit.

As in Glass et al.\ (1999), we have not applied extinction corrections to the
ISOGAL data. The absorption in the ISOGAL bands is believed to be less than
0.05 mag, though the precise character of the interstellar
extinction curve in this wavelength region is not well-determined.

\subsubsection{Astrometric calibration}

The extracted source positions were set systematically by reference 
to the Deep Near-Infrared Survey (DENIS) positions for the same area. 
The DENIS survey has an internal astrometric accuracy of order 0.5$''$ 
(see Omont et al.\ 1999) and is ultimately referred to the USNO-A2.0 
astrometric catalogue, which has a rms absolute accuracy of 1 arcsec. 
The rms dispersions of the differences between the ISOGAL and the 
DENIS positions are (0.6, 0.7) and (0.9, 0.8) arcsec in (R.A., Dec) 
for NGC\,6522 and Sgr\,I, respectively. 

\subsubsection{ISOGAL color-magnitude diagram}

Figure 1 shows the color-magnitude diagram (CMD) 
for 182 and 287 stars detected 
at both wavelengths in NGC\,6522 and Sgr~I fields, respectively.
The objects in the Baade's Windows are almost exclusively from the Bulge,
and are therefore at a nearly constant distance from the Sun, with 
a distribution governed by the thickness of the Bulge. The minimum 
photometric scatter due to line-of-sight effects can be expected to be 
similar to that derived from the period--$K$ luminosity plot for Miras 
in the Sgr I field (Glass et al.\ 1995), viz 0.35 mag, since the intrinsic
scatter in the relationship is known to be $\leq$ 0.13 mag from Magellanic 
Cloud studies (Glass et al.\ 1987). There is a continuous progression 
of [15] mag, and therefore dust output, with [7]--[15] color and 
late spectral type (Glass et al.\ 1999). 
The sequence stretches from the top of the Red Giant 
Branch (RGB), located in the bottom left corner of the diagram, to the 
Mira variables, which are generally, but not exclusively, the most 
luminous dust emitters. Mira-type large-amplitude variability is thus 
seen not to be a necessary condition for mass-loss in M-stars, as was 
previously believed.  

The objects which approach the Miras in dust emission are close to them 
in $K$-mag also (Frogel \& Whitford 1987; Glass et al.~1999), indicating 
that they have similar bolometric mags. Spectral types are available 
for all the M-giants in part of the ISOGAL NGC\,6522 field (Blanco 1986),
as well as for late-type M-giants in the whole field (Blanco, McCarthy 
\& Blanco 1984). It is evident from figure 12 of Glass et al.~(1999) 
that only M5 giants or later types are detectable at 15$\mu$m, i.e., it is 
only these objects that have observable dust shells. Mira and
SR variables of C-type are entirely absent from these fields. 

A group of stars with luminosities similar to Miras, marked by crosses in 
Figure 1, were examined by T.~Lloyd Evans on the photographic plate 
material that was used for finding most of the known Miras in the Baade's
Window fields (Lloyd Evans 1976). Little additional evidence was found 
for photometric variability, ruling out the notion that they may have 
been overlooked as Miras, but not excluding the possibility that they 
could be SR variables of much lower amplitude.  It therefore became of 
interest to see if a modern photoelectric survey would reveal anything more.

\subsection{MACHO}

The MACHO Project had dedicated use of the 50-inch Great Melbourne Telescope
located at the Mount Stromlo Observatory in Australia from January, 1992 
to January, 2000. A system of corrective optics installed at the prime 
focus gave a focal reduction to $f/3.9$ and a $1^{\circ}$ field of view.  
A dichroic beam-splitter enabled simultaneous blue and red imaging.
The MACHO filters were non-standard, with the
blue filter running from $\sim$4500$-$6300 $\AA$ 
and the red filter from $\sim$6300$-$7600 $\AA$
(see Alcock et al.~1999).  At both the red and blue foci, a mosaic of four
2048$\times$2048 Loral charge coupled devices (CCDs) were mounted, yielding 
an 0.52 square degree imaged area.

Approximately 45 square degrees of the Galactic bulge were observed every
few nights, with exceptions for weather and the southern summers.
Photometry was handled by a special purpose crowded-field, PSF-fitting 
code called SoDOPHOT, which is described by Alcock et al.~(1999).  
At the time of this work, the MACHO photometry database
contained a time-series of $\sim$1000 two-color photometric
measurements spanning $\sim$6 years for
most Galactic bulge fields.  Most of the stars of interest here
are quite bright, and thus the typical
error on each photometric measurement is about $\pm$0.02 mag.

Following Alcock et al.~(1999), the MACHO instrumental photometry have 
been calibrated to the standard Kron-Cousins system using:
\begin{equation}
V \ = \ v \ + \ 23.699 \ - \ 0.1804 \ (v-r)
\end{equation}
\begin{equation}
R \ = \ r \ + \ 23.412 \ + \ 0.1825 \ (v-r)
\end{equation}
where $v$ and $r$ are instrumental magnitudes, and $V$ and $R$ 
are on the Kron-Cousins standard system.  
The color coefficients are averages of the values determined for different 
CCDs in the MACHO focal plane, corrected for airmass.  
These calibration formulae are estimated to have an overall absolute 
accuracy of $\pm$0.10 mag in $V$ or $R$, and $\pm$0.04 mag in $(V-R)$.  
For the purposes of calibration, we assumed $(v-r)$~= 0.5~mag for 
all stars without instrumental colors in the database;
the systematic error may be larger for these stars. 
The systematic calibration error may also be larger for stars with colors 
of $(V-R)$~$>$~1.2~mag (see Alcock et al.~1999).

Astrometry for the MACHO database was derived independently for each field 
using the Guide Star Catalog. Astrometric offsets of order $\sim$1$''$ 
between MACHO fields (as determined from common stars in field overlap 
regions) are typically found, which gives an indication of the overall 
astrometric accuracy to be expected.

\section{Matching MACHO \& ISOGAL Sources}

In our initial reconnaisance of the MACHO photometry database,
we overplotted MACHO ``tiles'' on the spatial distribution of ISOGAL sources
(tiles are a defined region of the sky in the MACHO database,
each approximately a few arcminutes square).
In this manner, we identified 54 tiles that overlapped (or nearly
overlapped) ISOGAL sources.
Unique starlists were extracted
from the MACHO photometry database for these tiles, which totaled
just over $3\times10^{5}$ stars.  We estimate that 91\% and 88\% of
the areas of the ISOGAL SGR~I and NGC\,6522 fields, respectively, are
included in the MACHO photometry database.  
This estimate accounts for gaps in MACHO sky
coverage between fields and CCDs, but not for area lost to CCD defects
or saturated pixels.  The latter probably accounts for no more than
$\sim$5\% of the field area.

A blind spatial matching of the ISOGAL sources with
such a large number of MACHO stars would have
certainly resulted in numerous 
chance coincidences.  Therefore, we decided to apply
a ``reasonable'' cut in the optical color-magnitude diagram 
before attempting to cross-correlate sources.
We considered only MACHO stars with 
$V > 13.5 + 4.67(V-R)$ as possible counterparts to the ISOGAL
sources.  This cut was chosen to include the
red clump, very red faint stars, and very bright
blue stars.  
It excluded most faint main sequence stars, but
may have also excluded
some relatively blue AGB stars located behind the Galactic bulge.
After applying this cut, 
approximately $4\times10^{4}$ MACHO stars remained.


For each of the 1193 ISOGAL sources, the angular distance to each 
MACHO star was calculated, and the closest positional match was recorded 
if lying within 3 arcseconds. In an initial matching trial, median offsets 
in $\alpha$ and $\delta$ were calculated using a subset of matches that 
included only MACHO stars with colors $(V-R) > 1.5$ mag, which were 
considered very likely matches.  The offsets for each field, in the 
sense of ISO~$-$~MACHO, and in units of arcseconds, are: ($\Delta\alpha$, 
$\Delta\delta$) = (1.4, 0.7), and (0.8, 0.6) for ISOGAL fields SGR~I 
and NGC\,6522, respectively. A second and final matching trial proceeded 
as the first, except that the offsets were applied.

Figure 2 shows the final distribution of angular separations
for matches in ISOGAL fields SGR~I and NGC\,6522.
In order to provide some estimate of the
number of spurious matches, we repeated the 
matching procedure described above, except that we applied an
arbitrary ($\sim 15$ arcsec) shift to each MACHO starlist.
The result of this control matching trial is
also shown in Figure~2 (shaded histograms).  
The total number of matches are 518 and 386 
in ISOGAL fields SGR~I and NGC\,6522, while the probabilities
for a spurious match are 7\% and 3\%, respectively.
The details of the matching criteria
described above do not affect the main results of this work.

The total number of matches (904) represents 76\% of the 1193 ISOGAL 
sources, which compares to the estimated $\sim$90\% overlapping
area between the MACHO and ISOGAL fields.  This discrepancy 
($\sim$165 stars) is probably too large to attribute to CCD defects 
in the MACHO images, which would tend to lower the percentage of overlapping
area ($\sim$5\%).  It would also seem too large to attribute to saturation 
in the MACHO data ($V_{SAT} \sim$ 11; 
although this is certainly case for some known Miras, 
as discussed below).  It is unlikely that this represents a significant 
population of self-obscured AGB stars, since there aren't that many very 
bright ISOGAL sources.  It is possible that the ISOGAL sources without 
optical counterparts are not real, but are instead low signal-to-noise, 
false detections.  However, a combination of
all of the above effects is perhaps the most likely explanation.  A detailed
accounting of each ISOGAL source is beyond the scope of this work.  We are
confident that we have correctly matched a large number of ISOGAL sources 
with MACHO stars, and will proceed with an analysis of this dataset.

332 stars of the 904 MACHO and ISOGAL matches have complete sets of 
$V$, $R$, $[7]$, and $[15]$ mags.  They are listed in Table 1. In 
Figure~3, we plot all 904 sources, distinguishing those with four-color 
photometry. The four-color sources are seen to deviate in the direction 
of higher $V-R$ and/or fainter $V$ from the others, apart from the small 
group of stars near $V$ $\sim$ 17, $V-R$ $\sim$ 0.9, which coincide with 
the red giant clump. This can be an effect of either or both
of photospheric temperature and reddening due to circumstellar material. 
Those located near the clump are consistent with chance matches.

For the remainder of this work, we restrict our analyses to the
subset of 332 matches with four-color photometry.  

\section{Pulsation characteristics}

\subsection{Periodicity}

Most of our sample (305 out of 332, or 92\%) show quite well-defined 
periodic or quasi-periodic variations of moderate amplitude superimposed 
on irregular longer-term fluctuations, which do not, with a few 
exceptions, appear periodic on the overall time scale of our observations. 
Of the remaining 27 stars, 26 were frequently saturated in the MACHO image 
data and thus have unusable light curves. The remaining object is an 
eclipsing binary. There are 14 known Miras in these fields and they are 
discussed separately below. The SRs outnumber the Miras by about 20:1.

\subsection{Semiregular Variables}

All 332 $R$-band light curves were plotted and examined. Approximate 
periods were first estimated by eye and later made more precise by 
Fourier analysis. The seasonality of the observations led to the presence 
of confusing noise at low frequencies as well as aliases. Stars with 
estimated values of log $P$ $<$ 2.2 were analysed separately from those
with longer periods. 

The log $P$ $<$ 2.2 stars (280) were Fourier analysed season by 
season and their amplitude periodograms were then summed. 
The component with highest amplitude with log $P$ $<$ 2.2 was 
taken to be the most characteristic or relevant one, even if 
other periods sometimes had almost equal amplitudes.  

The remaining 25 of the 305 periodic or quasi-periodic variables, i.e.,
those with estimated log $P$ $\geq$ 2.2, were also Fourier analysed, 
but using the information derived by eye from the light curves to 
assist in the interpretation.

It is traditional to divide the SR variable giants, whether having 
M-, C-, or S-type spectra, into subtypes SRa and SRb, according 
to their variability characteristics. The SRa types show 
persistent periodicity in the range 35--1200 days, though 
with amplitudes less than Miras (defined to have minimum 
amplitude $>$2.5 mag at $V$) and variability in amplitude 
and light-curve shapes. The SRb stars have much more poorly 
expressed periodicity in the range (20--2300) days, with slow, 
irregular changes, or even periods of constancy. Usually, however,
a mean period can be assigned to them.

About 3/4 of our variables can be classified as SRa and most of the
remainder as SRb. However, this classification scheme can only be 
regarded as subjective. The SRb types occur predominantly among the 
longer-period objects.

Figure 4 (a) -- (d) shows examples of the different types of variables.

\subsection{Mira Variables}

Of the fourteen Miras known from previous work (Lloyd Evans, 1976; Glass 
et al.\ 1995), six have good to recognizable Mira light curves (Table 2). 
Five were completely missing from the overall list of 332 identifications 
because they were saturated in the MACHO template images for these fields 
and thus do not appear at all in the MACHO photometry database (Alcock et 
al.~1999). 
The remaining three were misidentified and {\it do} appear amongst the 332,
but as matches to nearby faint stars. The Miras themselves were saturated 
in the MACHO images. Two of the resulting three `wrong' light curves 
exhibit irregular flare-like spikes which presumably arise from occasional 
seeing-related contamination from the true Mira counterparts.

The periods determined by MACHO are based on much more comprehensive light
curves than either those of Lloyd Evans (1976) or Glass et al.\ (1995). It is
therefore interesting to compare these values in order to have some
check on the error arising from uncertain $P$ values when determining 
$P-L$ relations. The differences may be due to observational error 
or to intrinsic irregularities in the Mira light curves (Fig 4a). For the 6 
stars in common with Lloyd Evans (1975) we find $\Delta P / P_{\rm MACHO} 
= 7 \pm 5 \%$ and for the 4 stars in common with Glass et al.\ (1995) 
we get  $\Delta P / P_{\rm MACHO} = 9 \pm 9 \%$.  Since the scatter 
of the LMC Miras around their $K$, log$P$ relation is around 13\%, it is
clear that errors arising from uncertain or poorly determined periods 
contribute significantly to the overall scatter in the relationship, 
besides those associated with finding the mean values of $K$.  

\subsection{Period-Luminosity Relations for Semiregulars and Miras}

Fig 5a shows the log $P$, [7] diagram for our sample. As we will see, 
the 7\,$\mu$m mag is, like the $K$, closely related to $M_{\rm bol}$ 
for these stars. We have superimposed a line equivalent to 
that which fits local SR variables having photometric and astrometric 
data from Hipparcos, as suggested by Bedding and Zijlstra (1998).
To transform from their empirical $M_K$ relation to one involving [7], 
we made use of 51 late-type non-Mira stars in the NGC\,6522 Baade's Window 
having $K_0$ (de-reddened) values by Frogel \& Whitford (1987) and also 
7$\mu$m mags from ISOGAL, 
finding that [7] = 1.04 $K_0$ -- 0.20 (s.d. 0.26). 
(For reference, using Frogel \& Whitford's (1987) bolometric magnitudes 
corrected for the difference in assumed distance moduli, we also found  
$M_{\rm bol}$ = 0.75 [7]$_{\rm obs}$ -- 9.25 (s.d. = 0.21) for the same 
sample. The distance modulus of the Baade's Windows was taken to be 
14.7 in our work and 14.2 by Frogel \& Whitford). The Bedding and
Zijlstra (1998) line then has the form
\begin{equation}
[7]=-1.85\;{\rm log}P+11.27.
\end{equation}

The Bedding \& Zijlstra (1998) line runs fairly centrally through, or perhaps 
0.1 to 0.2 mag above, the distribution of Baade's Window points. It is 
about 0.8 mag above that originally found for globular cluster  
variables with periods in the range $ 0 \leq {\rm log} P \leq 2.8$ 
by Whitelock (1986). 

Fig 5a also includes an empirical log\,$P$, [7] fit for the Miras, excluding 
TLE 57 (a possible SR):
\begin{equation}
[7]=(-6.9\pm1.4)\;{\rm log}P+(23.5\pm3.4),
\end{equation}
with s.d. 0.4. Unlike the case for the SR variables, the 7\,$\mu$m flux 
from a Mira is likely to include a substantial component arising from dust 
as well as normal photospheric emission. It is therefore no longer a 
simple measure of bolometric output, particularly at longer periods. 

The $K$ emission of Miras is little contaminated by dust and 
{\it is} thus more representative of bolometric output. Accordingly, 
in order to facilitate a more direct comparison with other work, 
in Fig 5b we have transformed the [7] mags of the SRs to $K_0$ and 
plotted a log\,$P$, $M_K$ diagram, with the Bedding and Zijlstra 
(1998) line and the same authors' transformation of the Whitelock (1986) 
fit (its exact position is dependent on the distance scale used). 

Fig 5c shows the log$P$, $M_{\rm bol}$ diagram, where we have transformed 
the [7] mags of the SRs to $M_{\rm bol}$ using the relationship given
above. The Miras and their $P-L$ relation from Glass et al.\ (1995) are
shown, as is the (linear) $P$, $M_{\rm bol}$ relation given by Whitelock 
(1986) for Galactic globular clusters. 

The shallow sequence of SRs relative to the Miras in these diagrams 
(Figs 5a,b,c) may reflect an evolutionary sequence. Evolutionary tracks
of this kind, though covering a much reduced period range (1.8 $<$ log $P$
$<$ 2.8), were predicted by Vassiliadis \& Wood (1993). Alves et al.~(1998)
have also projected theoretical evolutionary tracks onto the 
PL diagram, extending the sequences to the lower luminosities 
and shorter periods appropriate for SRs. The Alves et al.~(1998)
PL sequences are based on accurate analytic approximations 
to the grid of Vassiliadis \& Wood (1993) AGB models and are thus 
properly comparable to the latter authors' PL sequences. 
The absolute luminosities of the evolutionary PL sequences depend on 
initial mass and metallicity. 
Alves et al.\ (1998) showed that SRs in the clusters 47~Tuc and NGC\,1783, 
which have similar metallicities but different initial masses, support 
the relative luminosities predicted by the theoretical PL sequences. 
Therefore, the overall luminosity of the SRs in Fig.\ 5a,b,c may indicate 
the characteristic age and metallicity of the AGB population.


When a star moves up along one of the nearly parallel evolutionary
sequences in the PL diagram, it eventually reaches the Mira line
at a unique position.  For Galactic                                          
globular clusters, this occurs at the relatively short period of about       
200 days, in accordance with the known period distribution of the Miras      
that they contain. The solar neighbourhood line intersects the Miras at      
about 460 days. The period distribution of local Miras is unknown at         
present, but long periods are common and this figure may be reasonable.      
The census of Miras in the SgrI Baade's Window is complete (Glass et al.\     
1995) and their average period is 346 days, in accord with the distributions  
in Fig.\ 5a,b,c.  This period is much longer than the typical 200 days 
found for Miras in Galactic globulars and is consistent with the SRs 
in Baade's Window lying above the Whitelock (1986) PL sequence.  

It is known that the scatter of the Miras in the Sgr I Baade's Window 
around the log\,$P$, $K_0$ relation is $\sim$0.35 mag (Glass et al.\ 1995). 
Most of this is attributed to the distribution of the Miras along the line 
of sight, i.e., the finite thickness of the Bulge, though some of it 
may be caused by the patchy nature of the interstellar extinction. The 
scatter of the SR variables in Fig.\ 5a is not much greater, implying that 
the spread of evolutionary tracks cannot be very wide. 
 
As can be seen from Fig.\ 6 (upper panel), there is no conspicuous clumping in
numbers at any given period of the semiregular variables in our sample.
There is however a gap visible in Fig.\ 5a,b,c between the Mira region 
and the semi-regular variables similar to that noted by Wood \& Sebo 
(1996) and Wood (2000) in the case of the Large Magellanic Cloud.
This gap corresponds, however, to the period range where aliasing is severe 
due to the seasonality of the MACHO data, and may in part be an artefact.  
Wood \& Sebo (1996) and Wood (2000) interpret the Mira sequence as being 
one of fundamental mode pulsators and suggest that the semi-regulars 
pulsate in higher modes. First overtone pulsators are expected to lie 
on a parallel sequence about $\Delta$log$P$ $\sim$ 0.35 
to the left of the fundamental, and higher modes should lie at progressively 
smaller intervals to the left of these. The sequences arising from the
higher modes, although clearly seen in the LMC, are not expected to be 
separated from each other as clearly in our Baade's windows data because of 
greater observational error and a range of distances. The necessity 
that the short-period end of the semi-regular M-giants must pulsate in 
very high modes has also been discussed by Koen \& Laney (2000). This 
interpretation is not incompatible with the evolutionary picture previously 
discussed.

\subsection{Amplitudes}

Because the short-period variability of the SRs is often modulated by
apparently irregular long-period trends, eye-estimates of the envelope of
the $R$-band variations were made for each star, accurate to about 20\% or 
0.05 mag, whichever is greater. None exceeded $\Delta R$ $\sim$ 1 mag, 
whereas the five Miras with adequate data (Fig.\ 4a) each showed 
$\Delta R$ $\sim$ 4 mag.

Three stars with short periods around 50 -- 60 days also showed long
periods around 400 days (one or two other short-period stars are also
suspected to have long periods close to a year, but their interpretation is
confused by the annual nature of the observations). They have 
amplitudes of 0.1 -- 0.2 mag in the short periods and 0.5 mag in the 
long (see Fig.\ 4d), but have luminosities appropriate to SRs rather than 
Miras. Similar stars are found in the LMC MACHO data (see
Alves et al.\ 1998; Wood et al.\ 1999), covering short periods of 50 -- 100 
days and long periods of 250 -- 1000 days. Wood et al.\ find long to short 
period ratios in the range 5 -- 13.
  
Figure 6 shows a histogram of the average amplitude of variability as well 
as the period distribution. There is a very clear trend towards smaller
amplitudes at shorter periods (see also Minniti et al.\ 1998). The 
shorter period groups may have been influenced by selection effects, 
in the sense that very small amplitude variables may not have been 
detected.

\section{Discussion of mass-loss rates}

In non-Mira M-giant stars, the flux entering the 7$\mu$m ISOCAM LW2 
filter arises primarily from photospheric emission and does not include 
a large dust contribution. The 15$\mu$m LW3 filter, on the other hand, 
overlaps with the silicate dust emission features at
10 and 18$\mu$ and is strongly 
affected by dust, when present. Only when the dust is optically thick, 
such as in long-period Miras and OH/IR sources, is the 7$\mu$m band 
likely to be strongly affected.   The $K_0$ and [7] mags of the SRs
and Miras previously discussed in \S4.4 are consistent with this
scenario.

\def\gtsima{$\, \buildrel > \over \sim \,$}
\def\ltsima{$\, \buildrel < \over \sim \,$}
\def\simgt{\lower.5ex\hbox{\gtsima}}
\def\simlt{\lower.5ex\hbox{\ltsima}}

\subsection{Spectral Energy Distributions}

Here we model the spectral energy distributions (SEDs) for the small 
subset of our sample also identified by Frogel \& Whitford (1987).  
These stars allow us to calibrate mass-loss rates based upon the 
observed 15 \micron\  flux excess using model SEDs.  We also
attempt an initial characterization of how mass-loss rates depend on 
fundamental stellar parameters such as luminosity and effective 
temperature in order to compare with recent theoretical predictions.  
This is the first time that dust sensitive mid-infrared
photometry has been assembled for a sample of SRs (and Miras) whose 
distances and thus energetics are known.  We compare pulsational
properties of the sample with mass-loss rates calibrated here in \S5.4.

First, we assembled $J_0$ and $K_0$ photometry (see also Table~2) and 
SED-integrated bolometric magnitudes for 26 stars (1 Mira from Table~1 
and 25 SRs from Table~2) from Frogel \& Whitford (1987).  The bolometric 
magnitudes were adjusted by $-$0.5 mag to account for a distance modulus 
of 14.7 mag as discussed previously. The ISOGAL running number (BW), the 
MACHO identifier, the star name by Frogel \& Whitford~(1987) and the 
adjusted values of $M_{\rm BOL}$ are listed in the first four columns of 
Table~3, respectively. The $V$ and $R$ magnitudes were dereddened by 
adopting a visual extinction of $A_V$ = 1.5 mag and taking $A_R/A_V$ = 0.75.
Optical and near-infrared fluxes were calculated using the zero points 
from Bessell, Castelli, \& Plez~(1998).  The mid-infrared fluxes were 
calculated from Eqns.~(1) \& (2) of this paper.  
Errors associated with absolute flux calibration 
and dereddening are negligible for the purposes of this work.

Stellar effective temperatures
in column~(5) of 
Table~3 were calculated from the 
dereddened $(V-K)_{0}$ color, where we employed 
the calibration of Bessell, Castelli, \& Plez~(1998).  
These temperatures are assumed to represent an accurate, relative
effective temperature scale.  It is noted that systematic 
uncertainty in the temperature scale at the
$\pm 100 ^{\circ}$K level or less will not affect our conclusions
(e.g., see \S5.2).

For our SED modeling, we assembled the ``corrected'' model spectra of 
Lejeune, Cuisinier, \& Buser (1997) with solar [Fe/H], $\log g$ of 
either 0.28 or 0.00, and effective temperatures of
2500, 2800, 3000, 3200, 3350, 3500, 3750, and 4000 $^{\circ}$K.  This was
the finest temperature grid available.  This
corrected spectral library, by definition, yields
synthetic optical and near-infrared broadband colors
consistent with the $(V-K)_{0}$
color-temperature calibration adopted above.
We caution that these model spectra are for static stars and do not take 
account of photospheric extension arising from variability (e.g., Bessell et 
al.~1989). 

The Lejeune et al.~(1997) model
spectra were input as $\lambda$, F$_{\nu}$ files into the
radiative transfer code, DUSTY (Ivezi\'{c}, Nenkova, \& Elitzur 1999).
DUSTY solves the radiative transfer problem of an AGB star 
enshrouded in dust, including a self-consistent solution for
the density structure in the wind-driven dust shell.  
We adopted the faster-running analytic approximation for the wind-driven 
dust density structure, which is an option in DUSTY.  We chose 100\% warm 
silicate for the grain composition  (Ivezi\'{c} et al.~1999).
The grain size distribution was a truncated power-law (q = $-$3.5, 
a$_{1}$ = 0.005 \micron\ and a$_{2}$ = 0.25 \micron; Ivezi\'{c} et 
al.~1999; see also Mathis, Rumpl, \& Nordsieck 1977).
The temperature at the dust shell's inner-boundary was fixed to be 
1000 $^{\circ}$K, which is supported by observations (Reid \& Menten 1997).  
We ran a total of $\sim$200 model SEDs with DUSTY using the 8 different 
Lejeune et al.~(1997) model spectra as inputs and allowing for a wide 
range of mass-loss rates.
DUSTY reports the mass-loss rate (\.{M}$_{\rm L4}$) and the
expansion velocity ($V_{exp}$) normalized
to a luminosity $L = 10^4 L_{\odot}$.  The true mass-loss rate (\.{M}) scales
in proportion to $L^{3/4}$ and $(r_{gd} \rho_{s})^{1/2}$, where $r_{gd}$ is
the gas-to-dust ratio and $\rho_{s}$ is the dust grain bulk density.  
The expansion velocity scales as $L^{1/4}$ and $(r_{gd} \rho_{s})^{1/2}$.
For these latter parameters, the default values from DUSTY are employed: 
$r_{gd}$ = 200 and $\rho_{s}$ = 3 g cm$^{-3}$.  This corresponds to an
absorption coefficient at 60~\micron\ of $\chi_{60}$ = 70 cm$^2$ g$^{-1}$.
For all stars, we assume that $r_{gd}$, $\rho_{s}$, the grain size 
distribution, the dust composition, and the temperature at the 
inner-boundary are the same. These simplifying assumptions are 
sufficient to establish a plausible, relative calibration of 
the mass-loss rates in our sample.

Each model SED from DUSTY is fit to the observed, dereddeded flux data in
the $\log(\lambda F_{\lambda})$ versus $\log(\lambda)$ plane, allowing for 
one zeropoint constraint.  The SED fits yield both the luminosity
and temperature.  
The best-fit input model
spectrum temperature (T$_{\rm MOD}$) and the
mass-loss rate normalized to $L = 10^4 L_{\odot}$ (\.{M}$_{\rm L4}$) 
are listed in columns (6) \& (7) of Table~3, respectively.
The true mass-loss rate (\.{M}), which is rescaled for each known 
stellar luminosity by a factor of $(L/10^4 L_{\odot})^{3/4}$, is 
listed in column~(8) of Table~3.
Finally, the expansion velocity corrected for a factor of
$(L/10^4 L_{\odot})^{1/4}$ is
listed in column~(9) of Table~3.

We find reasonable agreement between
the $(V-K)_{0}$ color temperatures and those indicated by the
best-fit model SEDs.  However, the former are preferred
because of the poor temperature resolution of the
grid of model spectra.  We also checked that
the stellar luminosities derived from our SED fits
agree with those taken
from Frogel \& Whitford (1987).  The agreement is fair in most
cases.  However,
in a few instances of fitting the SEDs of stars with 
high mass-loss rates, we obtain
luminosities lower than
those given by Frogel \& Whitford (1987).  In these cases, our statistical
best-fit model SED (which gives equal weight to all flux data) 
underestimates the 
near-infrared flux (near the peak of $\lambda F_{\lambda}$) and overestimates
the mid-infrared data.  Fortunately, while this 
affects the luminosities so obtained, 
the mass-loss rates themselves are not underestimated
at a level which is important in this work.
In summary,
we adopt the distance modulus-adjusted, SED-integrated
Frogel \& Whitford (1987) luminosities, the  
Bessell, Castelli, \& Plez~(1998) $(V-K)_{0}$ color temperatures, and
the SED-fit mass-loss rates in our subsequent analyses.

Figure 7 shows
four example SEDs (filled circles) and their best-fit DUSTY SEDs (solid 
lines).  Each input spectrum is shown as a
dotted line.
We plot $\log(\lambda F_{\lambda})$ versus
$\log(\lambda)$.

\subsection{Mass-loss rate, effective temperature, and luminosity}

Our data, as presented in Table 3, can be used to obtain observational 
constraints on the dependence of mass-loss rate on effective temperature 
and luminosity for the first time. We assumed that the relationship should
be of the form: \.{M} $\propto$ T$^{\alpha}$ L$^{\beta}$.
Adopting a factor of two uncertainty for each true mass-loss rate,
a chi-squared minimization yields the power-law exponents: 
\begin{equation}
\alpha \ = \ -8.80^{+0.96}_{-0.24} \;\;\; 
\beta \ = \ +1.74^{+0.16}_{-0.24} 
\end{equation}

Theoretical studies of mass-loss for C-rich long-period variables (not yet   
extended to O-rich or M-type variables) suggests that their mass-loss rates  
are mainly governed by stellar photospheric temperature, followed by mass    
and luminosity, and are relatively independent of amplitude of variation, C  
over-abundance, and pulsational period (Arndt, Fleischer \& Sedlmayr 1997). 
They obtain                                                                  
\begin{equation}                                                        
  \dot{M} \propto T^{-8.26}_{\rm eff} L^{1.53}.                                 
\end{equation}
which is in good agreement with our observations.

\subsection{Dust-based mass-loss rates from 15 $\mu$m photometry}

Table~3 includes the excess of 15 \micron\ flux       
based on a Rayleigh-Jeans extrapolation of the 7 \micron\ flux ($x$[15] in 
mJy) for each star. 
We can calibrate this quantity in terms of the true mass-loss rate, 
as given by the DUSTY models, yielding
\begin{equation}            
{\rm log} \dot{M} = 0.78(\pm0.08) \cdot {\rm log}(x[15]) - 7.88 (\pm 0.11),    
\end{equation} 
with a s.d. = 0.30 dex.                                                      
This formula should be valid for SRs having observed 15$\mu$m excesses 
from 3 to 130 mJy, giving mass-loss rates from 1 $\times$ 10$^{-8}$ to 5 
$\times$ 10$^{-7}$ $M_{\odot}$ yr$^{-1}$. 

As previously explained, determination of the 15$\mu$m continuum by 
Rayleigh-Jeans extrapolation of the 7$\mu$m flux will not be correct 
for objects whose dust shells become optically thick at shorter 
wavelengths than is usually the case for SRs. Thus, in particular, we do not 
expect long-period Miras to follow the empirical relationship given.  
         
It is instructive to compare our values with those given by the Jura (1987)
formula for the mass-loss from an AGB star:                       
\begin{equation}  
\dot{M} \ = \ 1.7 \times 10^{-7} \ \left(\frac{150}{\chi_{60}}\right) \
 v_{15} \ R^2_{kpc} \ L^{-1/2}_4 \ F_{\nu,60} \
\lambda^{1/2}_{10} \ \  M_{\odot} {\rm yr}^{-1},                                 
\end{equation}
where $v_{15}$ is the gas outflow velocity in units of 15 km s$^{-1}$,       
determined from CO observations, $R$ is the distance to the star in kpc,    
$L_4$ is the stellar luminosity in units of 10$^4$ $L_{\odot}$, $F_{\nu,60}$
is the flux from the object at 60\,$\mu$m in Jy, $\lambda_{10}$ is the 
mean wavelength of light emerging from the star in units of 10\,$\mu$m,
and $\chi_{60}$ is the dust absorption coefficient in units of cm$^2$ g$^{-1}$.
We adopt $R$ $\sim$   
8.7 kpc, $L_4$ = 0.3, ($M_{\rm bol}$ = --4.04) and $\lambda_{10}$ = 0.1 
from the bolometric magnitude of a 200-day Mira (Glass et al.\ 1995).
To relate the given 15\,$\mu$m flux to the 60\,$\mu$m flux  
required, we take the relation by Jura (1986), 
intended for carbon stars
(but see also the values of $Q_{abs}$ for astronomical     
silicate grains; Draine \& Lee 1984), namely $F_{\nu}$ $\propto$          
$\nu^{1.54}$. 
As noted above, our DUSTY models employ $\chi_{60}$ = 70 cm$^2$ g$^{-1}$.
Finally, we adopt the mean expansion velocity of those listed
in Table~3, or 16 km s$^{-1}$; this compares to
8 km sec$^{-1}$, the average value determined for         
semi-regular variables by Kerschbaum, Olofsson \& Hron (1996). 
If the excess 15\,$\mu$m flux is 100 mJy, we obtain           
\.{M} = 2 $\times$ 10$^{-7}$ $M_{\odot}$ yr$^{-1}$.  This is within
a factor of two of the mass-loss rate expected from our 
calibration, 4 $\times$ 10$^{-7}$ $M_{\odot}$ yr$^{-1}$ for
a similar 15$\mu$m flux excess.  Thus our mass-loss rate calibration 
and Jura's well-known calibration yield similar results.
           
\subsection{Mass-loss rate and Pulsation Periods}

The part of the 15\,$\mu$m flux emitted by dust ($x$[15]) for the whole 
sample of SRs has been estimated by subtracting the photospheric flux 
which was taken to be a Rayleigh-Jeans tail fitted to the 7\,$\mu$m 
measurements. The latter were assumed to be completely free of dust 
emission. 
Figure 8 shows this quantity $x$[15] as a function of log period. 
Although the same procedure 
has been applied to the Miras, which are also included in Fig.\ 8, it 
becomes unreliable with increasing period because the assumption that 
the 7\,$\mu$m flux is free of a dust component progressively ceases 
to be valid.                                          
                                                                           
It is evident from Fig.\ 8 that the mass-loss rates of the longer-period       
SRs overlap those of the short-period Miras and clearly do not depend on  
the presence of large-amplitude pulsation. The lack of measurable mass-loss  
for stars with periods $P$ $<$ 60 days accords with the finding of           
Kerschbaum, Olofsson \& Hron (1996, see below) that CO radio emission is not 
detected for 0 $<$ $P$ $<$ 75 days. Similarly, though stars with periods     
in the range 75 $<$ $P$ $<$ 175 days may show dust emission, as they may     
also show CO emission, this does not constitute a {\it sufficient}           
condition.                                                                   

\subsection{CO-based information}

As noted above, the period-dependent behavior of the quantity
$x$[15] for SRs in the Galactic bulge 
is consistent with that of CO detections
for SRs found in the solar neighborhood.
It is therefore of interest to briefly recall the
CO-based information. 
 
Kahane \& Jura (1994) found CO emission from 11 SRs with measured periods
(typically 100-160 days) and brighter than $K$ = 0. They determined
mass-loss rates of 1--1.5 $\times$ 10$^{-7}$ $M_{\odot}$ yr$^{-1}$, 
which they compared to calculations based on the dust mass-loss 
rates derived from IRAS 60$\mu$ fluxes. Similar dust-to-gas ratios, 
CO line ratios, outflow velocities and mass-loss rates were found 
as for Miras with 300 $\leq$ P $\leq$ 400 days period, leading to 
the speculation that this group of SRs were overtone pulsators
corresponding to the Miras which pulsate in the fundamental mode.
Kerschbaum, Olofsson and Hron (1996) extended their sample to another 
group of SRas and SRbs, selected on the basis of their 60$\mu$m fluxes. 
The Kerschbaum \& Hron (1992) blue SRVs were not detected, whereas the red
ones and the Miras had a 50\% detection rate. Those with periods 
below 75 days were not seen, nor were those with 175 $\leq$ P 
$\leq$ 325 days. Those with 75 $\leq$ P $\leq$ 175 had a high 
detection rate. A CO study by Young (1995) of nearby Miras 
with optically thin dust shells showed that only stars of type later 
than M5.5 could be detected, the rate becoming 100\% only for M7 and 
later types. The mass-loss rate was found to be correlated with
far-IR luminosity but not color, and also with CO outflow velocity. 
Similar mass-loss rates were found for M6.5--M8 types (comparable 
to the rates for semiregular variables found by other workers). 

\subsection{Mass-loss rates as a function of luminosity only}

We have seen that mass-loss in SRs can be expressed as a function of 
$T$ and $L$. High luminosities and long periods in LPVs are associated 
with low temperatures, which are not independent variables but are
connected by evolutionary tracks. 

Figure 9 shows $M_{\rm bol}$ plotted against the mass-loss rates 
from Table 3 of our work together with those derived for dust-enshrouded O- 
and C-rich AGB stars in the LMC by van Loon et al.~(1999).
There is a striking continuity between the rates exhibited by 
the low-luminosity semi-regular variables and those found amongst
the extreme AGB-tip variables, both C-type and M-type, in the LMC, in 
spite of the differences in metallicities between the samples.
We have derived a linear fit between log $\dot{M}$ and $M_{\rm bol}$. Because
the distribution of errors between these quantities is uncertain, we
have solved first assuming that all the errors are in $M_{\rm bol}$ and
second assuming that they are in $\dot{M}$. Both fits are shown and the 
average slope is given by the solid line, whose form is
\begin{equation}
\dot{M} \propto -1.09M_{\rm bol},
\end{equation}
i.e., we find that
\begin{equation}  
\dot{M} \propto L^{2.7},                                                   
\end{equation} 
in the range 10$^{-8}$  $<$ $\dot{M}$ $<$ 10$^{-4}$ $M_{\odot}$ yr$^{-1}$. 
                                                                           
\section{Conclusions}

1. Almost all non-Miras in our sample detected in the four MACHO and ISO
colors show semi-regular variability.

2. The SRs outnumber Miras by 20:1 in our sample.

3. We see no preferred periods in 10-200 day range, but a gap exists 
between the distributions of SRs and Miras which may be explicable in 
terms of pulsation modes.

4. The Galactic bulge SRs possess a $P-L$ distribution similar to 
that of the solar neighbourhood SRs observed by Hipparcos. 
They can probably be regarded as lying on a series of $P-L$ relations
with slopes equal to that observed in globular clusters, but with
luminosity levels appropriate to higher metallicities and initial masses. 

5. The amplitudes of the SRs increase with period, reaching about 0.3 mag
at 100 days.

6. Mass loss depends on luminosity and period, but does not require large 
Mira-like amplitudes, even though the mass-loss levels reach those of the
shorter-period Miras.

7. A minimum period of about 70d is required for, but does not guarantee, 
detectable mass loss, in agreement with conclusions based on CO 
observations.

8. The mass-loss rate for semi-regular variables depends on temperature and
luminosity approximately according to $\dot{M} \propto T^{-8.8}L^{1.7}$.

9. The observed mass-loss rates in SRs range from 1 $\times$
10$^{-8}$ $M_{\odot}$ yr$^{-1}$ to 5 $\times$ $10^{-7}$ $M_{\odot}$ 
yr$^{-1}$.

10. Taking into account the work of van Loon et al.~(1999) concerning extreme
mass-losing AGB stars in the LMC, we find the general result that $\dot{M} 
\propto L^{2.7}$ in the range 10$^{-8}$  $<$ $\dot{M}$ $<$ 10$^{-4}$ 
$M_{\odot}$ yr$^{-1}$. 
                                                                           
Note that in this work we have discussed only those stars which were detected
in both MACHO and both ISOGAL bands. A preliminary examination of the light
curves of the stars seen by ISOGAL only at 7\,$\mu$m indicates that most of
them are also SRs, but presumably with mass-loss rates too low for
15\,$\mu$m detection.

\begin{acknowledgements}

D.R.A. acknowledges support of this work
from a NASA grant administered by
the American Astronomical Society.  D.R.A. also thanks the South 
Africa Astronomical Observatory for his appointment to the Visiting 
Astronomer Program, and acknowledges their 
financial support of his visit.  

I.S.G. thanks the Institute of Astronomy, University of Cambridge, and
the Institut d'Astrophysique, Paris, for their hospitality and 
support during part of this work, under PPARC and CNRS grants, respectively.

We thank M.~Groenwegen for useful comments on an early version of this paper.

The MACHO Collaboration thanks the 
skilled support by the technical staff at MSSSO.
Work at LLNL was performed under the auspices of the U.S. Department 
of Energy bu the University of California Lawrence Livermore National \
Laboratory under contract No.\ W7405-ENG-48.  Work at CfPA 
was supported by NSF AST-8809616 and AST-9120005.  Work at MSSSO was supported
by the Australian Dept.~of Industry, Technology and Regional Development.
W.J.S. thanks PPARC Advanced Fellowship, K.G. thanks DOE OJI, Sloan, and
Cottrell awards, C.W.S. thanks Sloan and Seaver Foundations.
D.M. is supported by Fondecyt 1990440.

\end{acknowledgements}

\clearpage
\begin{figure}
\plotone{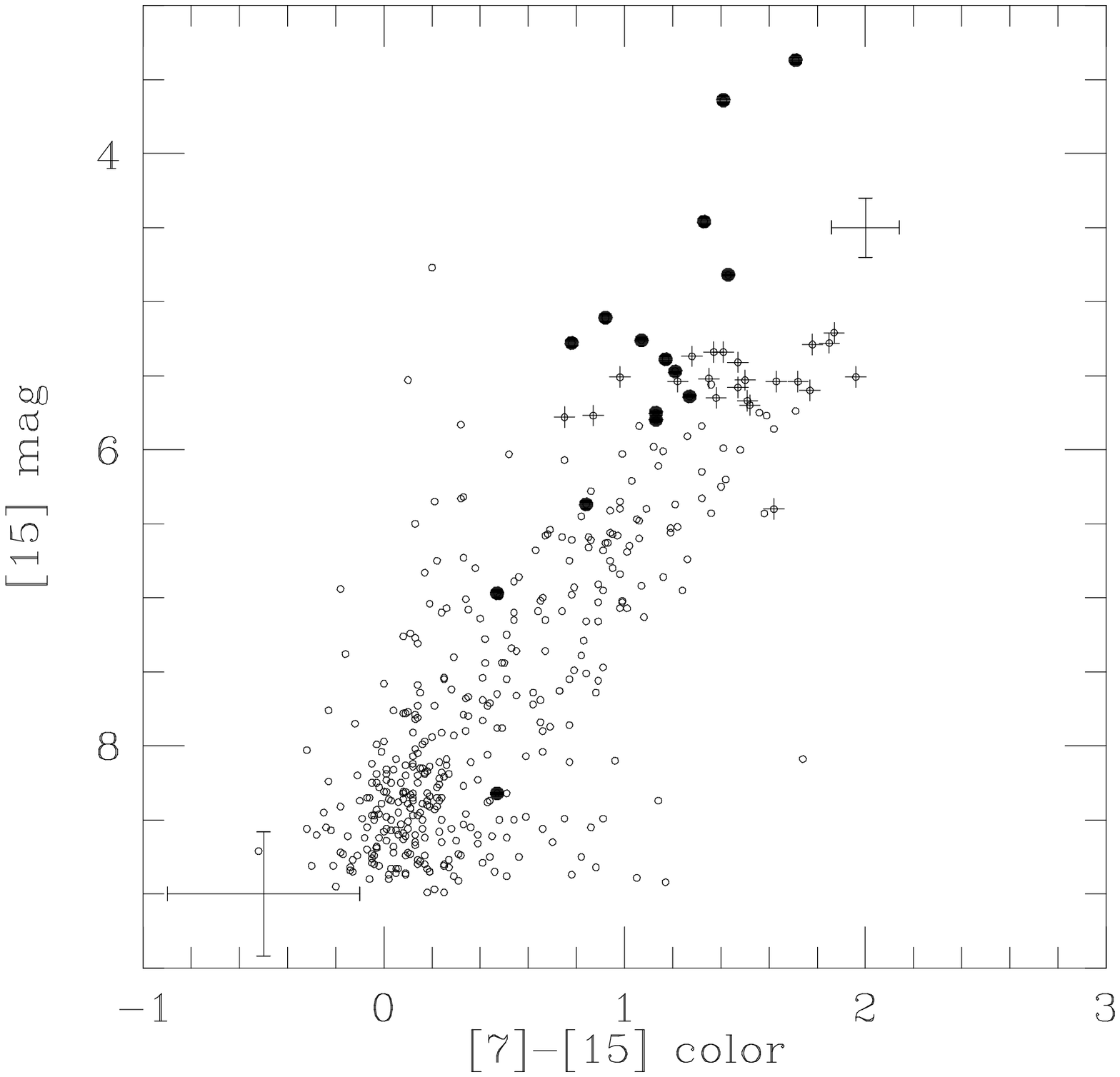}
\caption{Combined ISOCAM color-magnitude diagram for the 15 $\times$ 15 
arcmin$^2$ fields in the NGC\,6522 and Sgr\,I windows. Note the 
characteristic sequence of increasing 15\,$\mu$m flux, representative 
of mass-loss, with [7] -- [15] color. The heavy dots represent Mira 
variables and the crosses are bright stars 
that were examined unsuccessfully for variability in the pre-MACHO data. 
The top of the red giant branch (RGB) is around [15] $\sim$ 8. The 
objects brighter than this level are asymptotic-giant-branch (AGB) stars, 
except for a few foreground objects. Representative error bars are shown 
at each end of the sequence.}
\end{figure}

\clearpage
\begin{figure}
\plotone{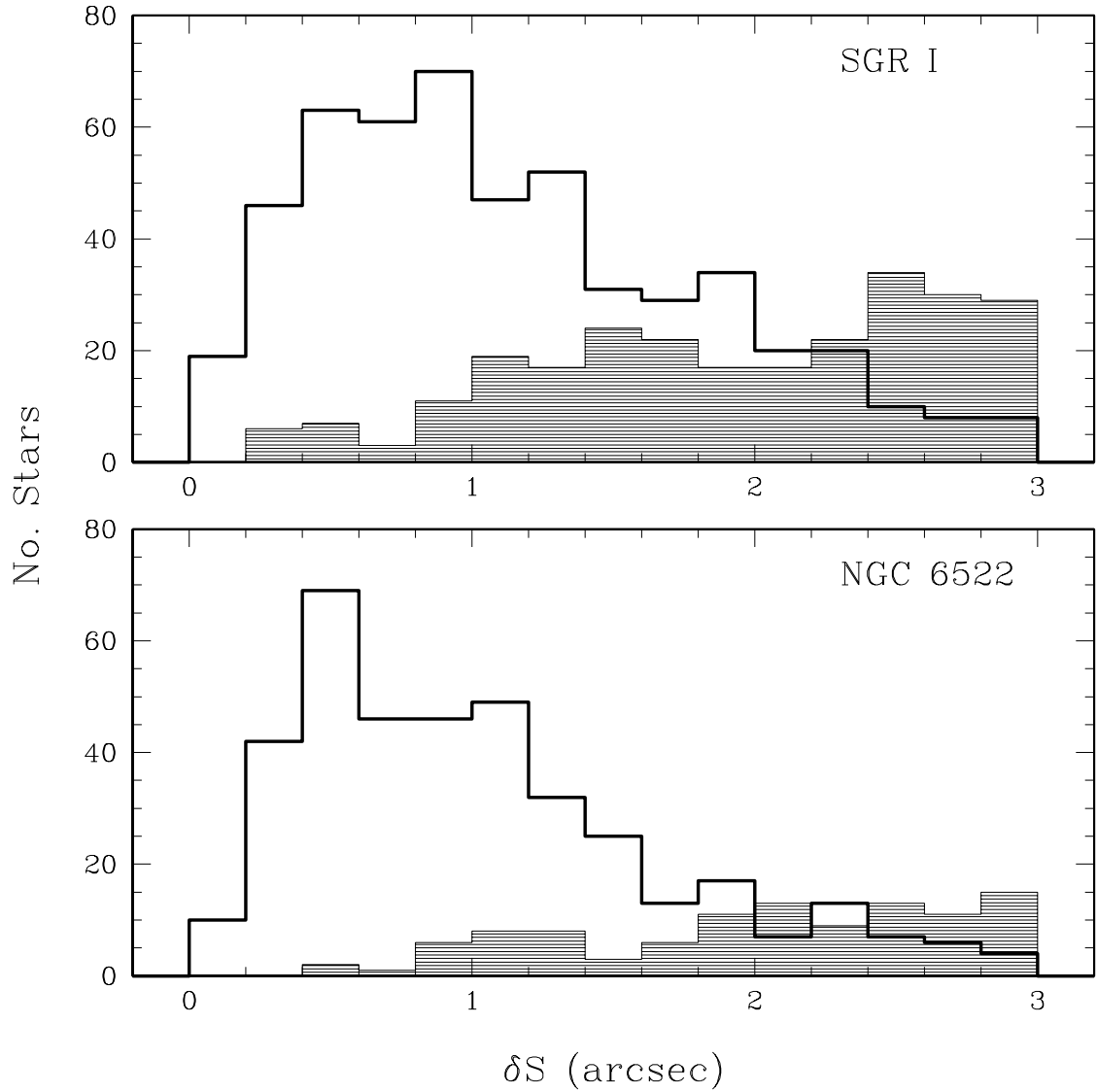}
\caption{The distribution of $\delta S$ (the angular distance between ISOGAL
and MACHO source coordinates in arcseconds) for the final match lists.
Example distributions of spurious matches are also shown (shaded histograms).}
\end{figure}

\clearpage
\begin{figure}
\plotone{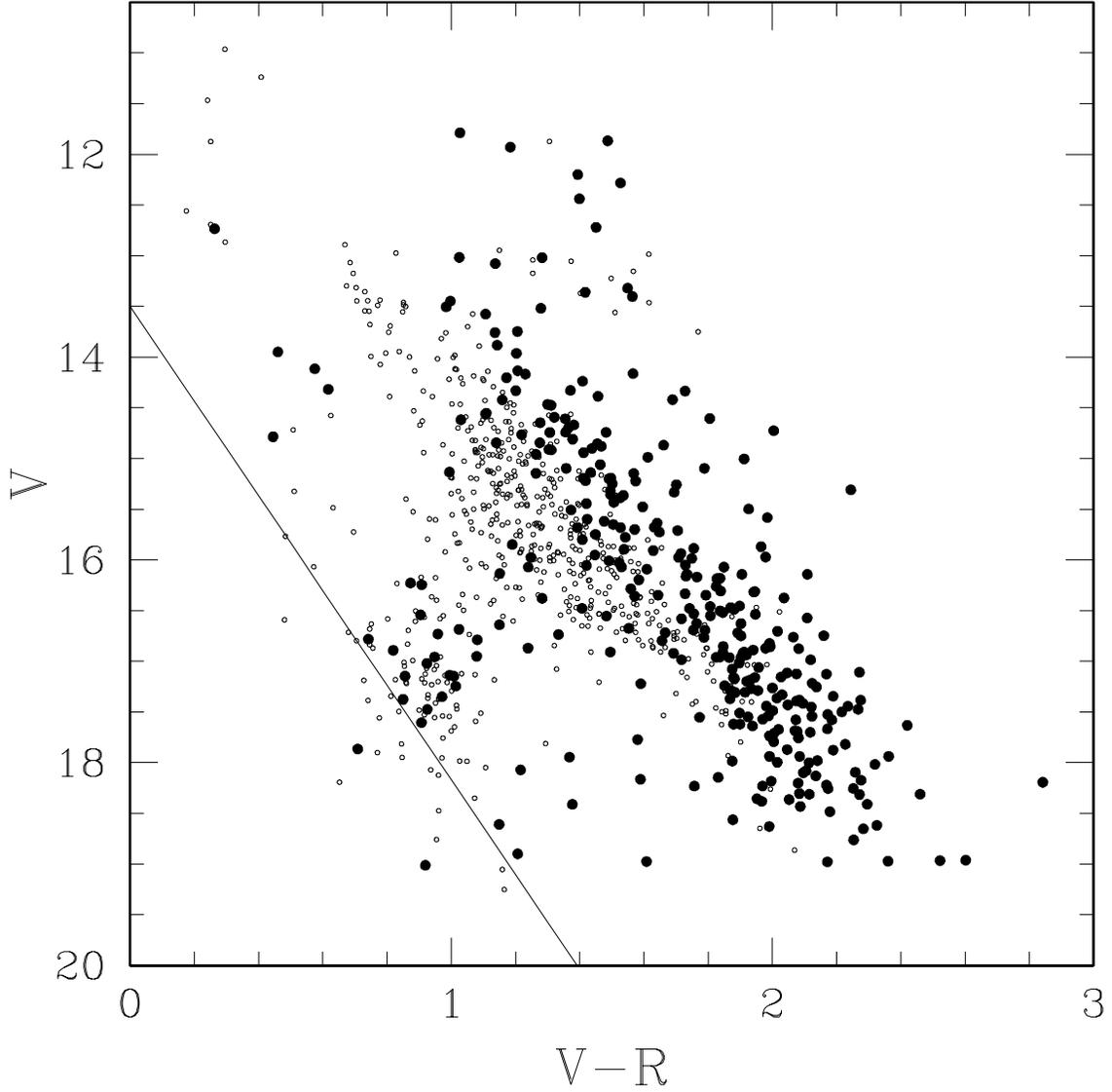}
\caption{The optical CMD for MACHO+ISOGAL sources in
the SGR~I and NGC\,6522 fields.  904 matched
sources with MACHO $V$ and $R$ colors are shown
as open circles.  Sources with four colors ($V$, $R$,
$[7]$, and $[15]$), are shown with bold filled circles.
The $V$,$(V-R)$ cut described in \S3 is shown as a solid
line.  The four-color sources lie mostly in the reddest
areas of the AGB.  Some sources are foreground stars, and 
some are possible spurious matches 
(i.e., near the red clump at $V$ $\sim$ 17, $V-R$ $\sim$ 0.9).}
\end{figure}

\clearpage
\begin{figure}
\plotone{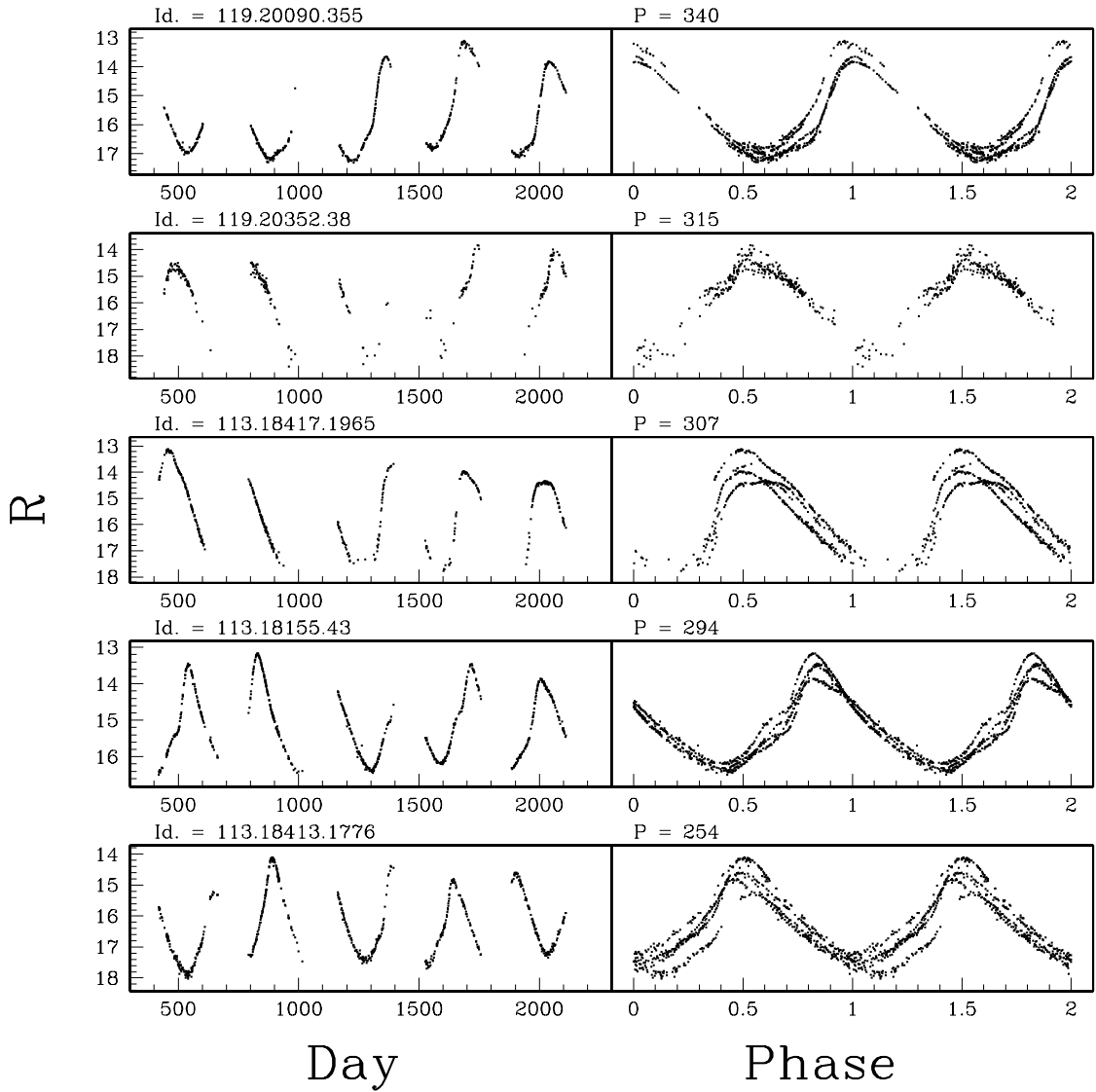}
\caption{(a) Left: MACHO $R$ light curves (mag vs.\,day)
for five Miras. Right: Same, folded 
according to period (mag vs.\,phase).}
\end{figure}

\setcounter{figure}{3}

\clearpage
\begin{figure}
\plotone{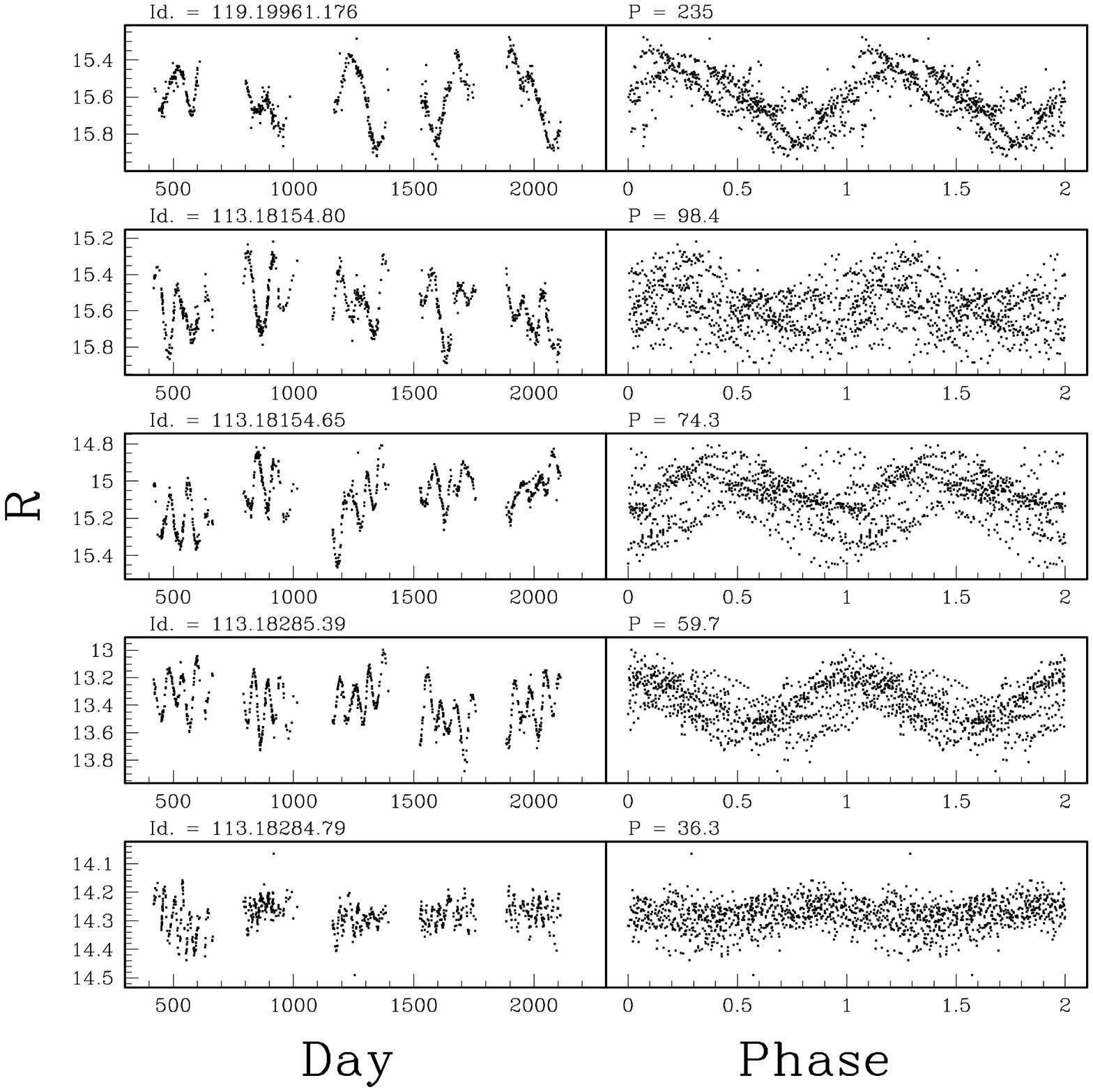}
\caption{(b) Left: MACHO $R$ light curves (mag vs.\,day)
for five variables classified SRa. 
Right: Same, folded according to period (mag vs.\,phase).}
\end{figure}

\setcounter{figure}{3}

\clearpage
\begin{figure}
\plotone{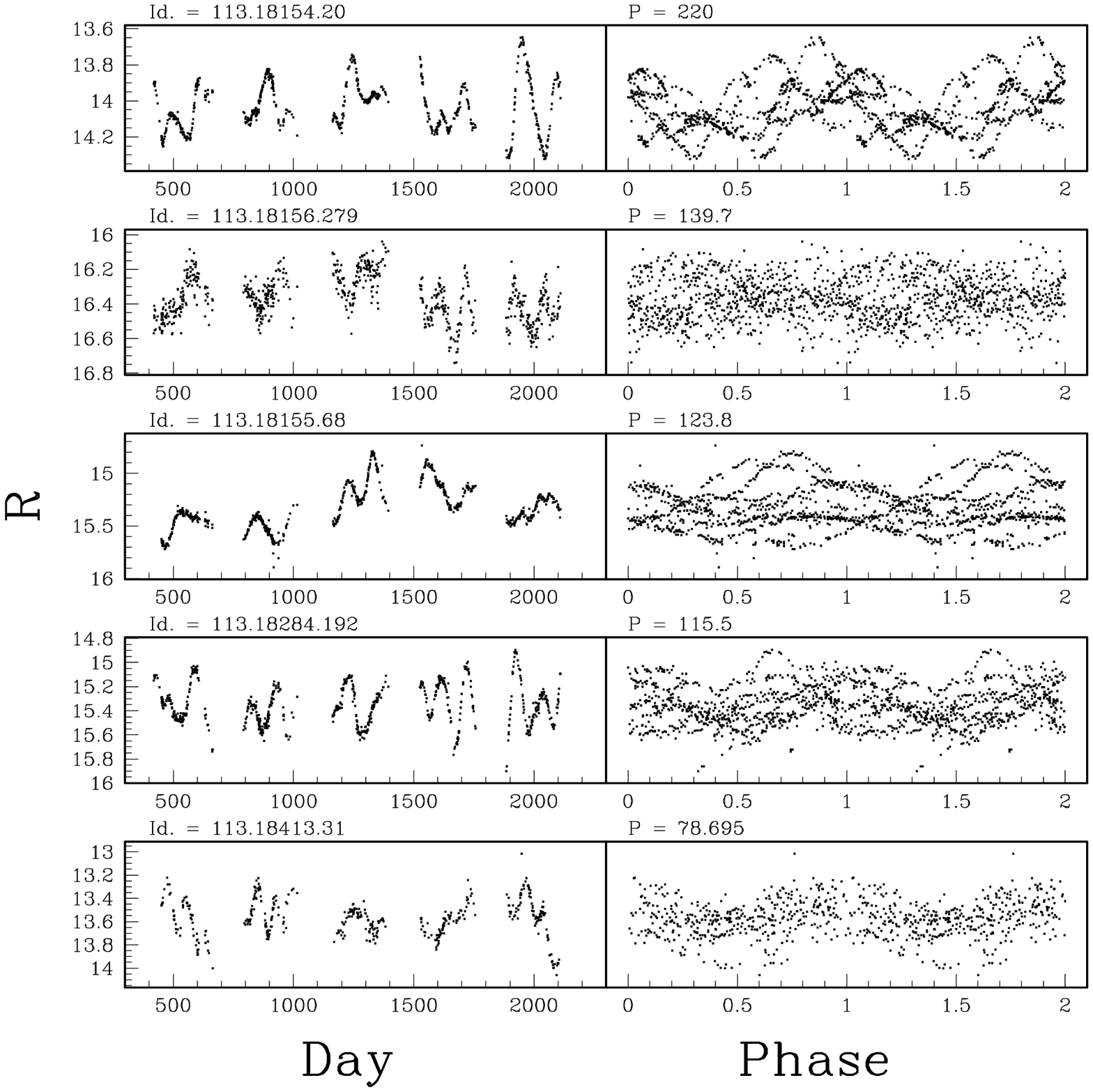}
\caption{(c) Left: MACHO $R$ light curves (mag vs.\,day)
for five variables classified SRb. 
Right: same, folded according to period (mag vs.\,phase).}
\end{figure}

\setcounter{figure}{3}

\clearpage
\begin{figure}
\plotone{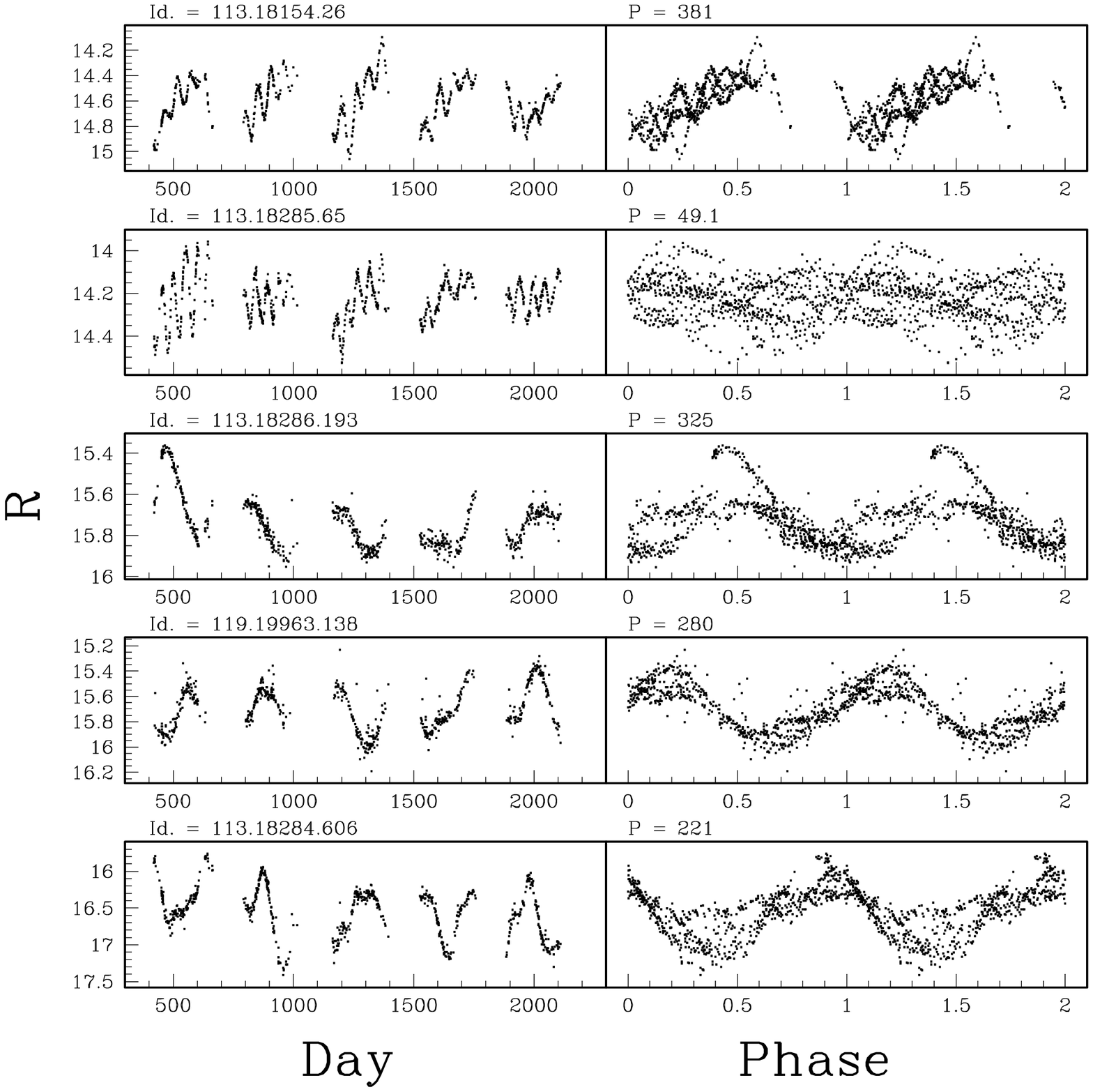}
\caption{(d) Left: MACHO $R$ light curves (mag vs.\,day)
of two double-period variables 
(upper panels) and three large-amplitude SRas (bottom panels). Right: same, 
folded according to period shown (mag vs.\,phase).}
\end{figure}

\clearpage
\begin{figure}
\plotone{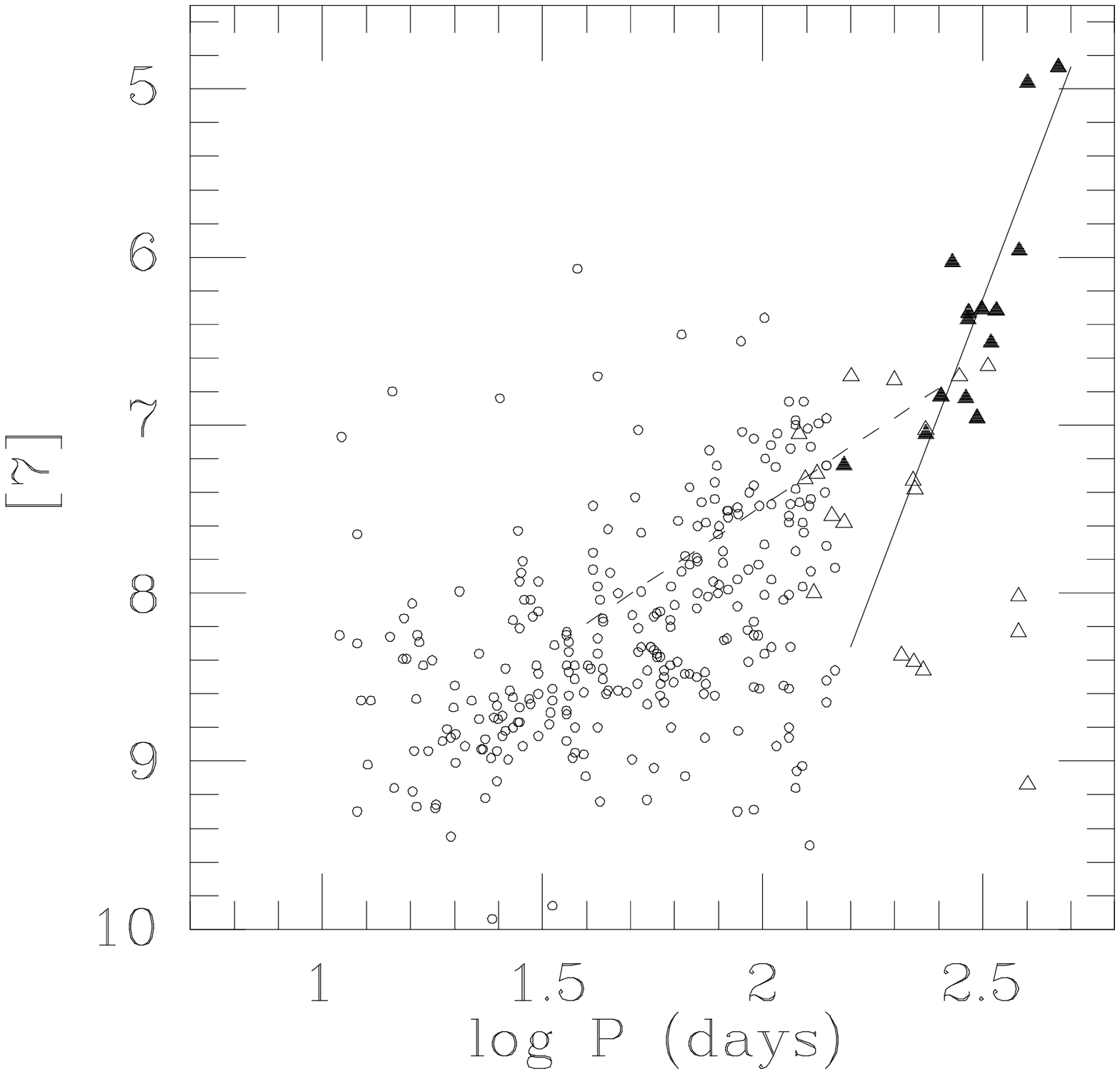}
\caption{(a) The [7] - log\,$P$ diagram ([7] in mags, P in days)
for cross-identified objects.
The open circles are the main periods identified. Open
triangles are separately identified long periods and solid triangles are
Mira (large-amplitude) variables, some of which were saturated in the MACHO
data. Periods for these cases were taken from Glass et al.\ (1995). The dashed
line is an empirical period-luminosity relation derived for semiregular
variables in the solar neighbourhood from Hipparcos and other data by
Bedding \& Zijlstra (1998). The solid line is an empirical fit to the Mira
data (see text).} 
\end{figure}

\setcounter{figure}{4}

\clearpage                                                                
\begin{figure}                                                            
\plotone{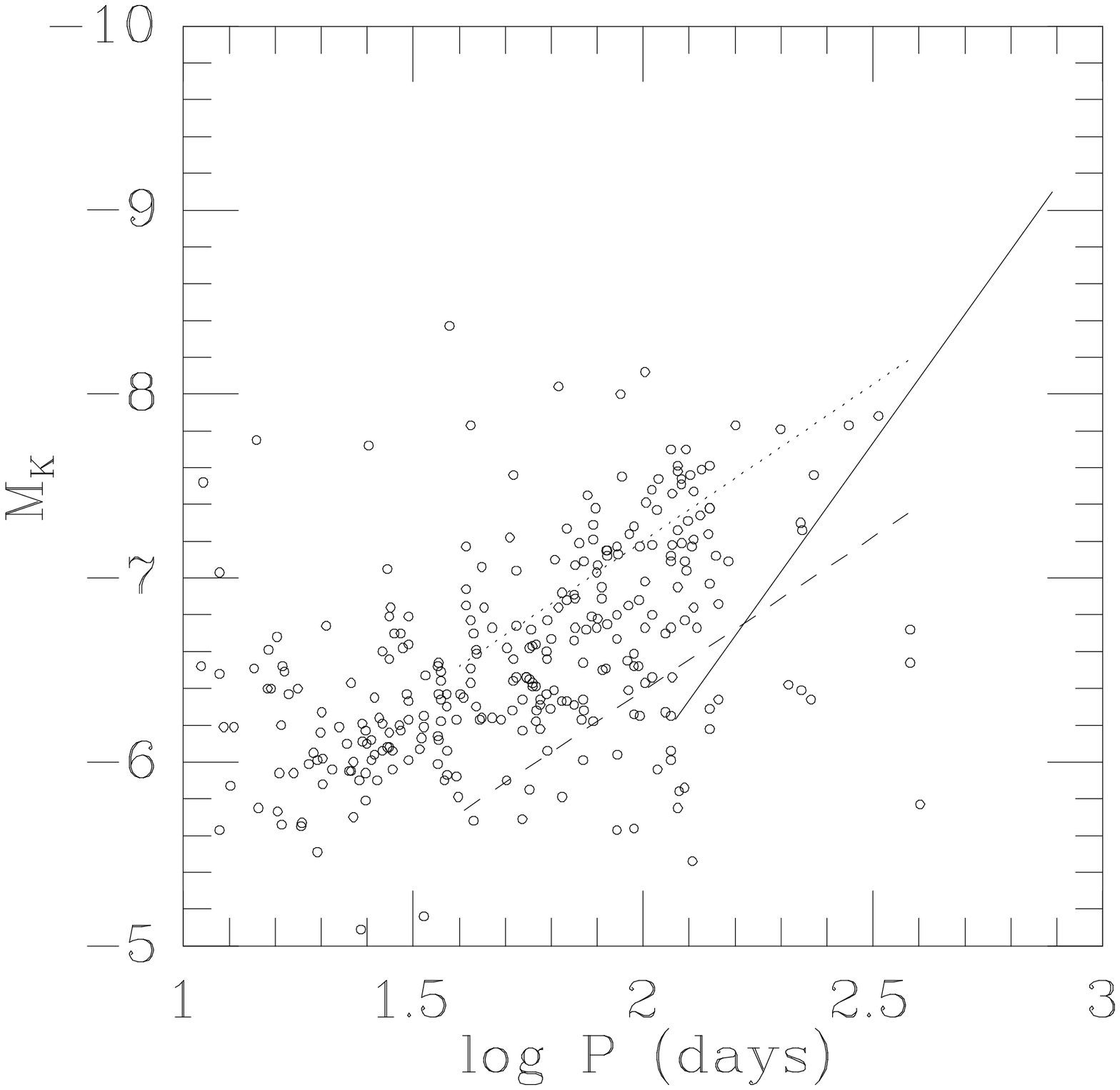}                                                  
\caption{(b) Similar to Fig 5a but for $K$, log$P$. 
The solid line is a fit to Mira data for the whole Sgr\,I field,
taken from Glass et al.\ (1995). (See Fig.~5c for the distribution
of Miras in a similar version of this diagram.)
The dotted line is from Bedding \& Zijlstra (1998), 
representing semi-regular variables in the solar neighbourhood with known 
periods and distances from Hipparcos.  
The dashed line is the relation for globular cluster variables from       
Whitelock (1986) as transformed by Bedding \& Zijlstra (1998).}                 
\end{figure}                                                              

\setcounter{figure}{4}                                                                          

\clearpage                                                                
\begin{figure}                                                            
\plotone{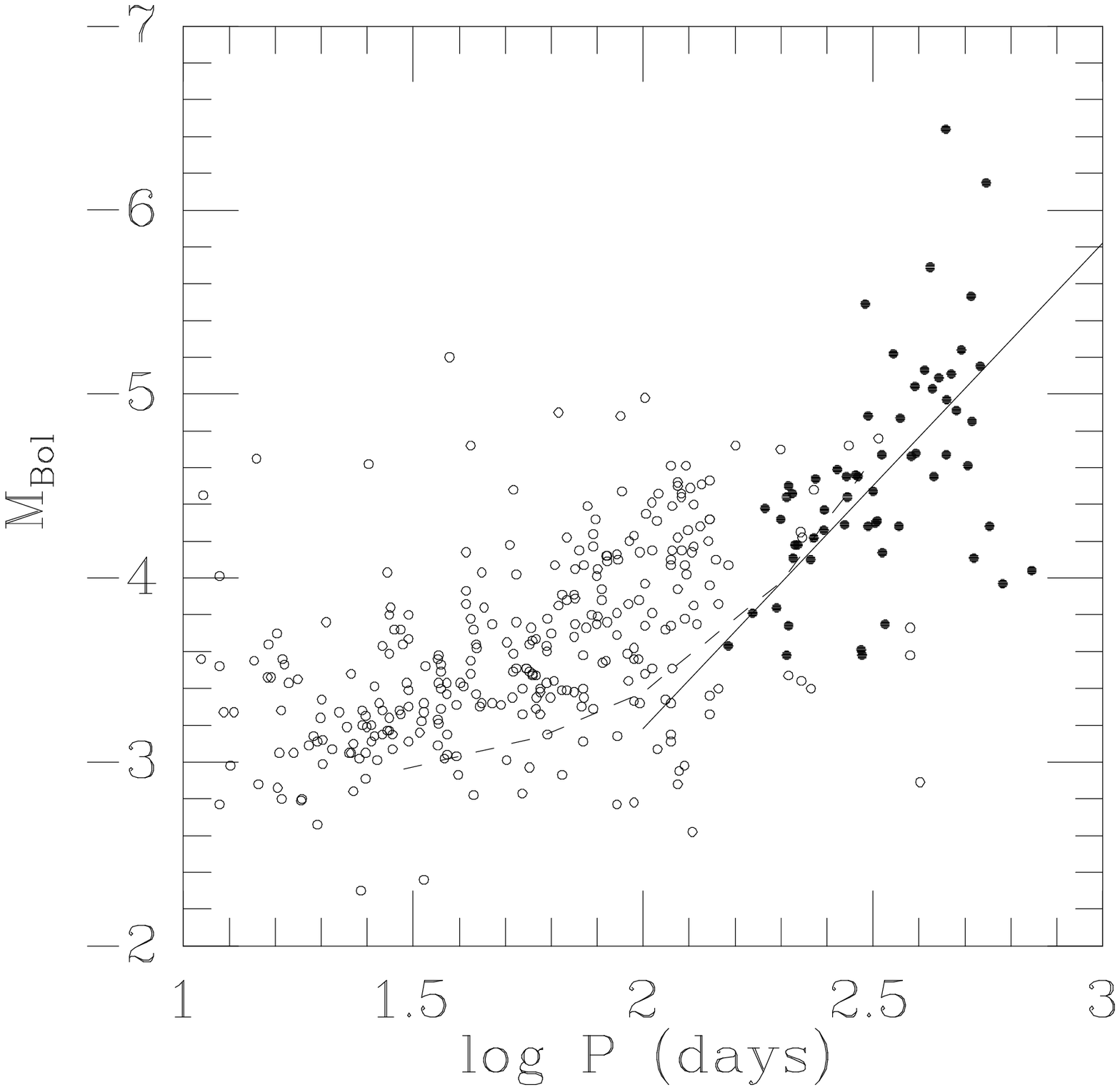}                                                  
\caption{(c) Similar to Fig 5b but for $M_{\rm bol}$, log$P$. 
Miras (dark dots) and their $P$--$L$ relation (solid line) for the
entire Sgr\,I field, have been taken from Glass et al.\ (1995). 
The dashed line is the
observed relation for galactic globular clusters, taken from
Whitelock (1986).} 
\end{figure}

\clearpage
\begin{figure}
\plotone{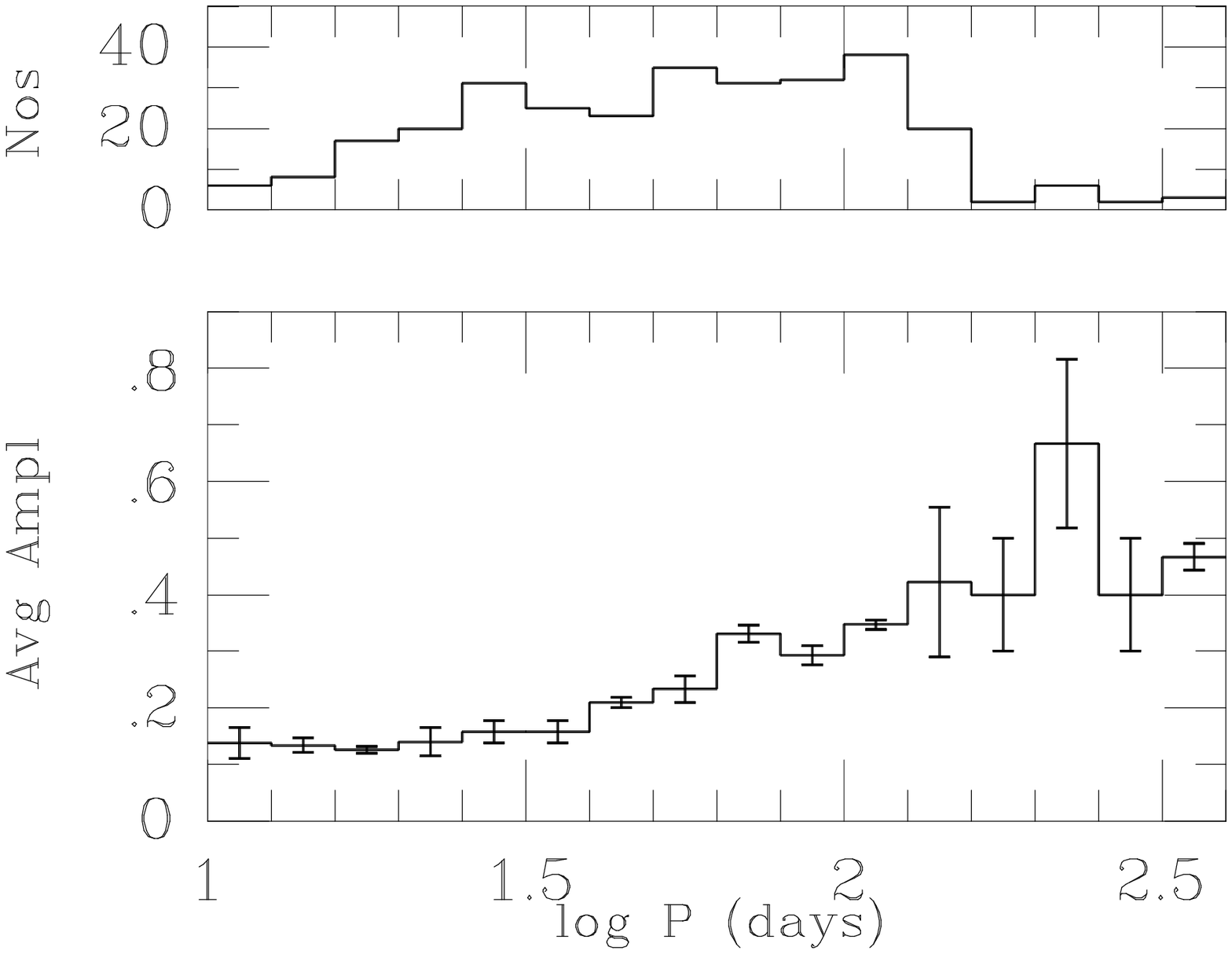}
\caption{{\it Lower: }Histogram of mean amplitudes of all variables for 
log(period) groups, excluding the Miras. Standard deviations are also 
shown.  
{\it Upper: }Numbers of variables in each log $P$ box.} 
\end{figure}

\clearpage
\begin{figure}
\plotone{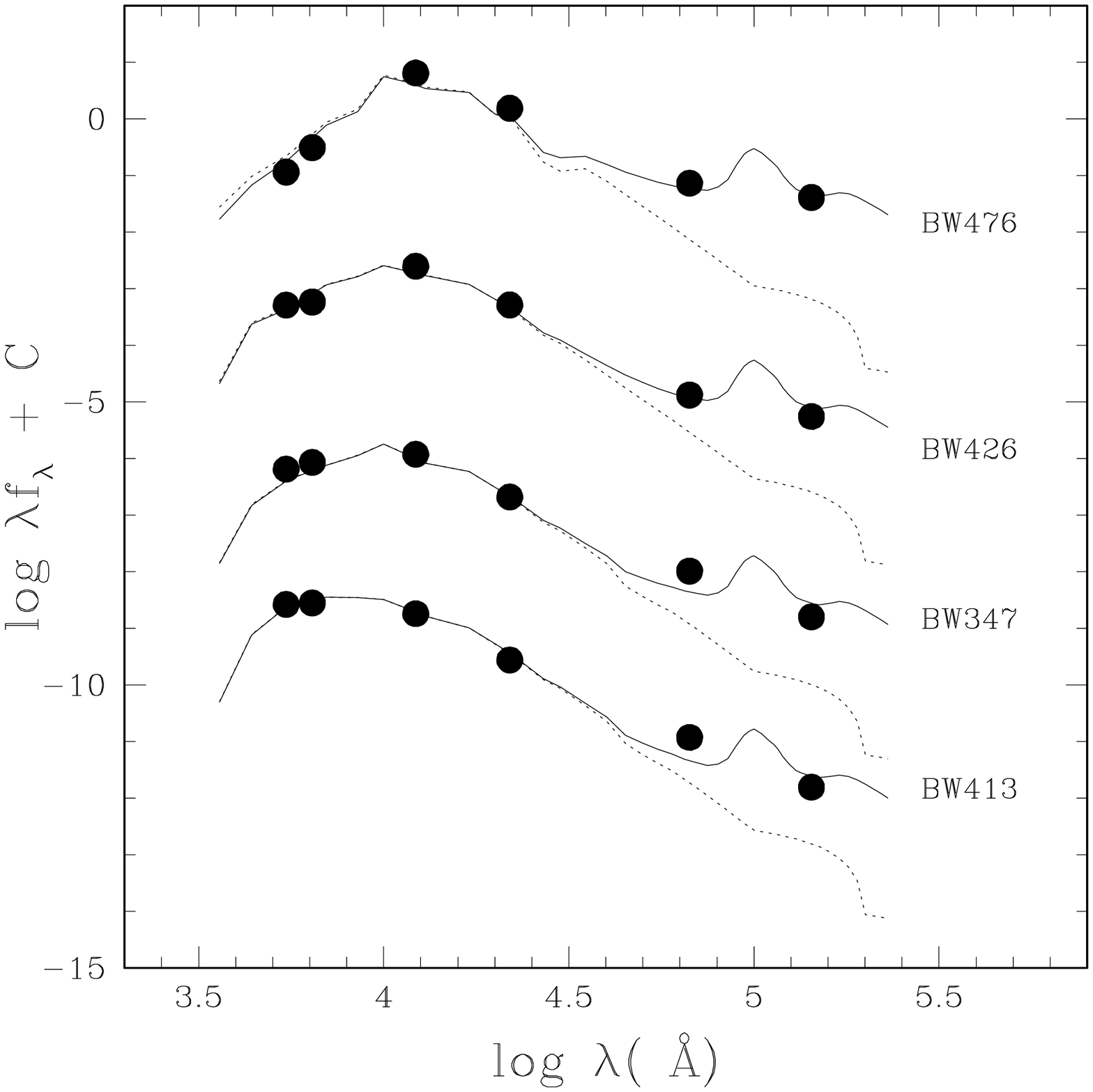}
\caption{Examples of observational data for four stars (filled 
circles) and best-fit model spectral energy distributions
(SEDs; $\lambda F_{\lambda}$ in ergs s$^{-1}$ cm$^{-2}$ and $\lambda$ in $\AA$)
including circumstellar dust
(solid lines; modeled with ``DUSTY''). Each input spectrum 
without circumstellar dust
is shown as a dotted line (same units as the DUSTY model SEDs).
The zero-point (C) is arbitrary and there 
is a vertical offset of 3 dex between each star.} 
\end{figure}

\clearpage
\begin{figure}
\plotone{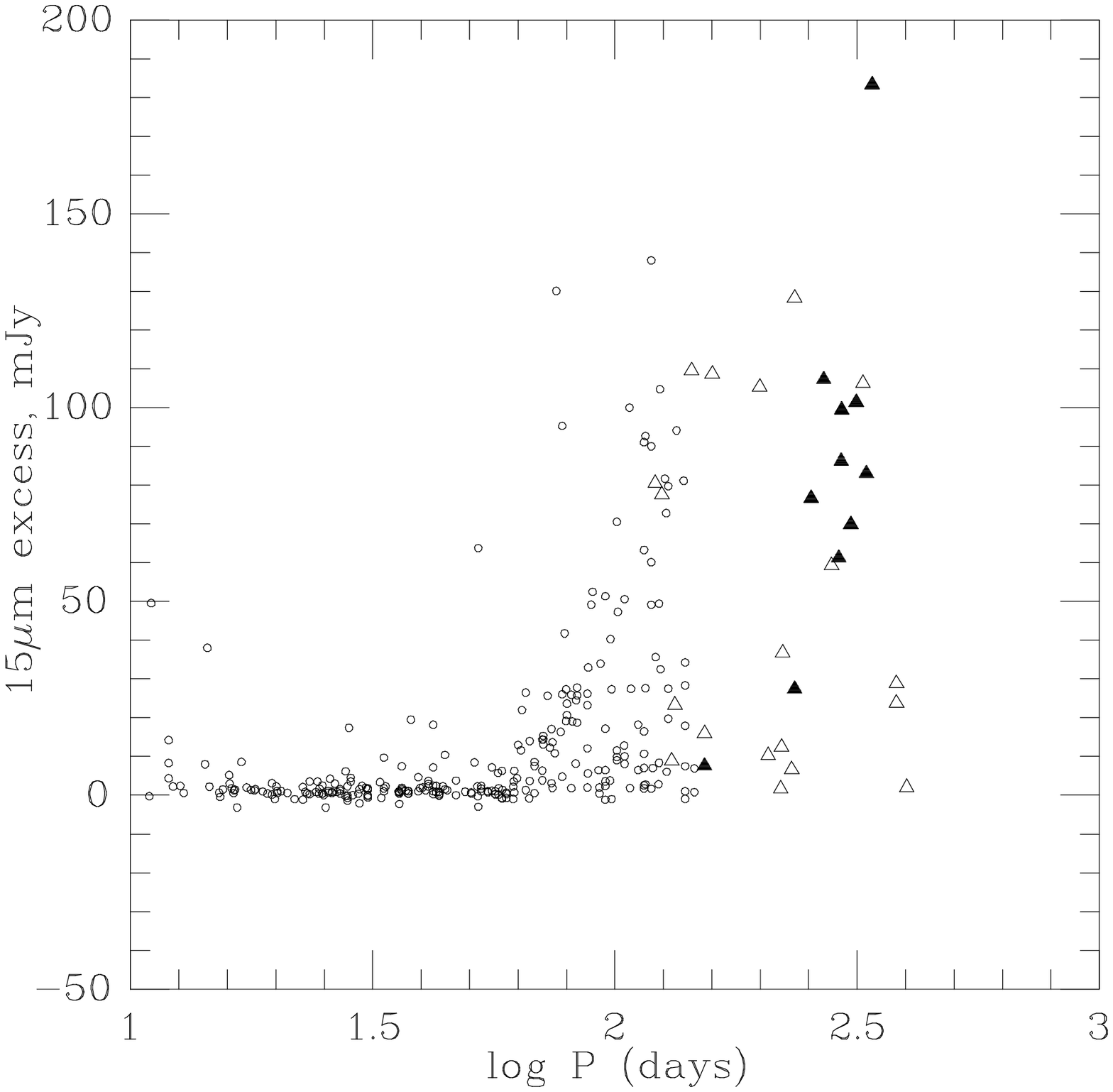}
\caption{Excess 15\,$\mu$m flux, an indication of mass-loss, beyond what is
expected by assuming a Rayleigh-Jeans photospheric energy distribution
fitted to the 7\,$\mu$m fluxes, shown plotted against log $P$. 
Symbols are as in Fig.~5a.  Having a
period $P$ $>$ 70 days appears to be a necessary, but not a sufficient
condition, for significant mass-loss.} 
\end{figure}

\clearpage
\begin{figure}
\plotone{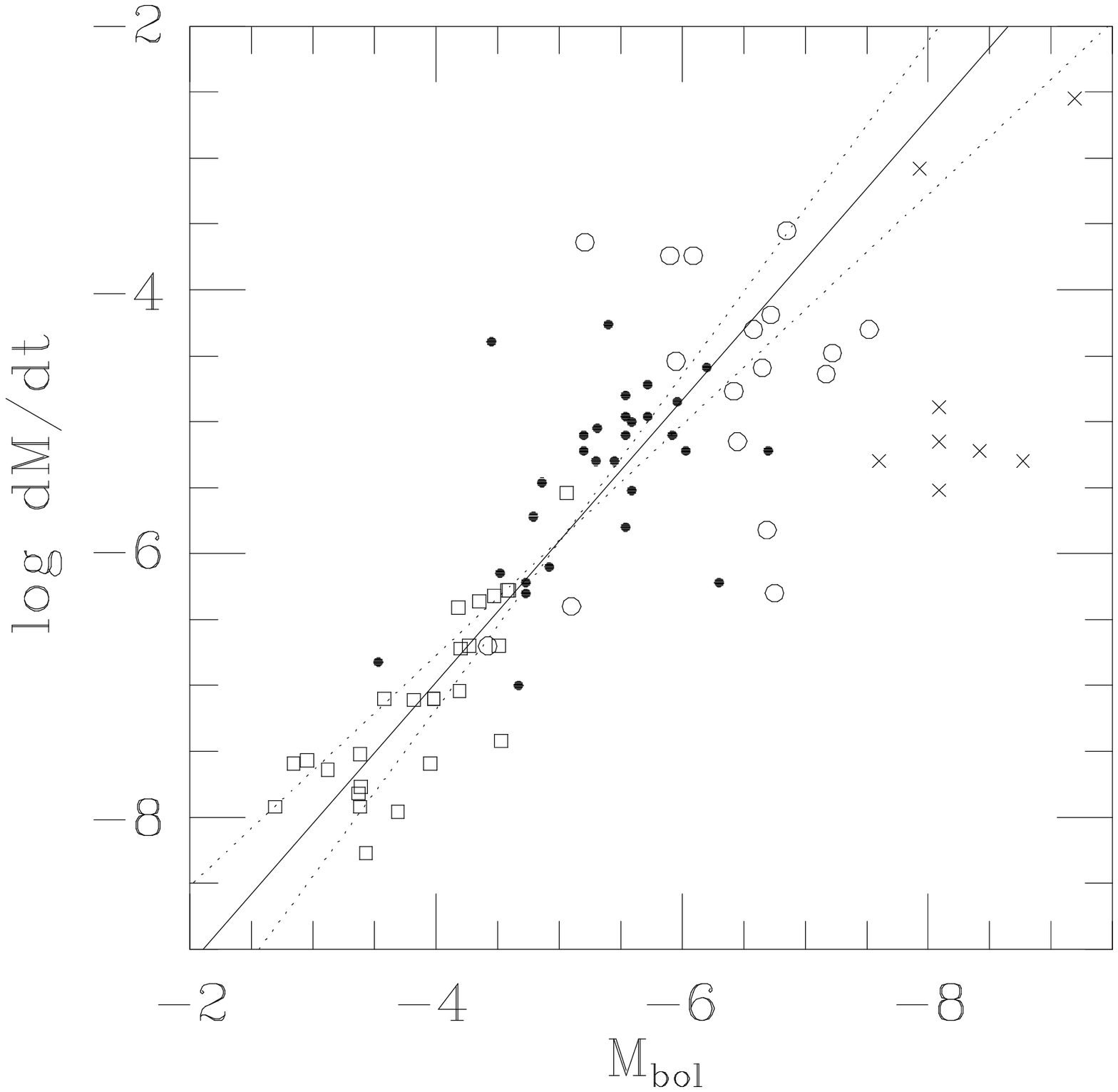}
\caption{$\dot{M}$ in units of $M_{\odot}$ yr$^{-1}$ plotted against 
$M_{\rm bol}$ for our sample of Baade's Window SRs (open squares) and
a sample of extreme mass-losing AGB stars in the LMC, 
from van Loon et al.\ (1999). The LMC C-type stars are shown as solid dots and
M-type stars as open circles. The crosses are M supergiants. The solid line
is a linear fit and the dotted lines give an indication of the errors in
the slope (see text).}
\label{fig9}
\end{figure}

\clearpage
\textheight22cm
\textwidth17cm
\hoffset-0.6in
\tighten
 
\begin{landscape}
\begin{deluxetable}{lllllrrrrlll}
\tablewidth22cm
\tablecaption{Summarized Photometry of Sources}
\tablehead{
\colhead{ISOGALP\tablenotemark{A} } &
\colhead{BW\tablenotemark{B}} &
\colhead{MACHO\tablenotemark{C}} &
\colhead{$V$\tablenotemark{D}} &
\colhead{$R$\tablenotemark{D}} &
\colhead{$J_0$} &
\colhead{$K_0$} &
\colhead{[7]} &
\colhead{[15]} &
\colhead{Amp.\ } &
\colhead{$\log P$\tablenotemark{E}} &
\colhead{notes\tablenotemark{F}} 
}
\startdata
J175840.5$-$290348& 54&113 18155   17&14.92 &13.47 &   &   &8.61&8.07&0.2 &1.891& \nl                                  
J175840.9$-$290827& 57&113 18154   52&16.81 &15.16 &   &   &8.46&8.37&0.2 &1.738& \nl                                  
J175841.8$-$290352& 61&113 18155   68&17.49 &15.38 &   &   &6.86&5.41&0.3 &2.093& \nl                                  
J175842.1$-$290650& 65&113 18155   18&15.90 &13.92 &   &   &7.12&6.73&0.25&2.019& \nl                                  
J175842.4$-$290515& 67&113 18155  251&19.06 &16.65 &   &   &8.04&7.07&0.4 &2.048& \nl                                  
J175842.6$-$290240& 68&113 18156   31&15.72 &14.01 &   &   &7.93&7.82&0.15&1.490& \nl                                  
J175842.6$-$291029& 69&113 18154  179&18.30 &16.10 &   &   &7.44&6.59&0.3 &1.891& \nl                                  
J175842.7$-$290340& 70&113 18156  252&18.69 &16.31 &   &   &8.59&8.07&0.25&1.594& \nl                                  
J175842.9$-$290845& 72&113 18154   65&17.25 &15.11 &   &   &7.58&7.00&0.3 &1.871& \nl                                  
J175843.4$-$291012& 74&113 18154   51&17.09 &15.21 &   &   &8.62&8.56&0.1 &1.433& \nl                                  
J175843.9$-$290325& 77&113 18156   21&14.87 &13.41 &   &   &8.58&8.59&0.2 &1.672& \nl                                  
J175843.9$-$290712& 78&113 18155   43&17.65 &15.69 &   &   &6.33&5.28&4.0 &2.468:& Mira  \nl                         
J175844.6$-$290234& 82&113 18156   38&15.68 &14.05 &   &   &8.55&8.28&0.12&1.301&   \nl                                
J175845.6$-$290357& 87&113 18155  118&17.64 &15.75 &   &   &8.29&8.17&0.15&1.560&  \nl                                 
J175845.9$-$290326& 88&113 18156   48&16.17 &14.43 &   &   &8.80&8.64&0.12&1.574&  \nl                                 
J175846.0$-$291033& 89&113 18154   84&17.32 &15.43 &   &   &8.95&8.87&0.1 &1.574&  \nl                                 
J175846.0$-$290748& 90&113 18154  108&17.68 &15.65 &   &   &8.67&8.61&0.15&1.396&  \nl                                 
J175846.2$-$290310& 93&113 18156   75&17.15 &15.08 &   &   &8.14&8.12&0.35&1.753&  \nl                                 
J175846.9$-$290722& 99&113 18155  462&18.08 &16.87 &   &   &8.11&7.96&0.1 &1.490&  \nl                                 
J175846.9$-$290332&100&113 18156   76&16.99 &15.10 &   &   &8.60&8.68&0.15&1.490&  \nl                                 
J175847.2$-$290711&104&113 18155  159&18.20 &16.02 &   &   &7.83&7.35&0.1 &1.834& \nl                                  
J175847.5$-$290157&108&113 18156   54&16.47 &14.66 &   &   &8.54&8.37&0.15&1.768&  \nl                                 
J175847.8$-$290943&111&113 18154   23&15.78 &14.24 &   &   &9.29&8.79&0.1 &1.980&  \nl                                 
J175848.0$-$291003&112&113 18154   20&16.17 &14.06 &   &   &7.33&7.26&0.3 &2.342:& \nl                               
J175848.2$-$290127&113&113 18156  279&18.74 &16.39 &   &   &7.72&6.52&0.3 &2.145&  \nl                                 
J175848.7$-$290743&117&113 18154   80&17.98 &15.60 &   &   &7.48&6.58&0.3 &1.993&  \nl                                 
J175849.2$-$291120&121&113 18154   26&16.53 &14.64 &   &   &8.02&6.74&0.5 &2.581:& dp(0.15,1.756)   \nl    
J175849.3$-$290528&124&113 18285  146&16.97 &15.14 &   &   &8.68&8.85&0.1 &1.298&  \nl                                 
J175849.5$-$291012&126&113 18284  606&18.98 &16.65 &   &   &8.41&7.47&1.0 &2.344:& \nl                               
J175850.2$-$290633&128&113 18285   55&14.93 &13.61 &   &   &8.52&8.40&0.1 &2.145&  \nl                                 
J175850.3$-$290455&129&113 18285  115&17.45 &15.15 &   &   &7.47&6.57&0.5 &2.063&  \nl                                 
J175850.6$-$291015&134&113 18284  413&18.14 &16.01 &   &   &7.64&6.53&0.5 &2.094&  \nl                                 
J175850.7$-$290314&135&113 18286   81&16.66 &14.73 &   &   &8.14&7.94&0.15&1.478&  \nl                                 
J175850.9$-$290106&136&113 18286  615&18.61 &16.70 &   &   &7.54&5.51&0.35&2.158:& \nl                               
J175851.2$-$290721&139&113 18285   35&15.02 &13.40 &   &   &7.60&6.99&0.4 &1.852&  \nl                                 
J175851.2$-$290206&140&113 18286  592&17.25 &16.24 &   &   &8.64&8.50&0.1 &1.523&  \nl                                 
J175851.6$-$290315&141&113 18286   54&15.69 &14.15 &   &   &8.15&8.20&0.08&1.185&  \nl                                 
J175852.0$-$290524&143&113 18285  523&17.15 &16.14 &   &   &8.17&7.16&0.1 &1.980&  \nl                                 
J175853.3$-$290422&154&113 18285  314&17.69 &15.74 &   &   &8.57&8.35&0.3 &1.523&  \nl                                 
J175853.3$-$290831&155&113 18284   79&15.92 &14.28 &   &   &8.61&7.86&0.1 &1.560& \nl                                  
J175853.6$-$285938&157&113 18287 1926&16.55 &14.59 &   &   &7.93&7.99&0.15&1.448&  \nl                                 
J175854.2$-$290351&162&113 18285   61&15.70 &14.13 &   &   &8.74&8.50&0.1 &1.390&  \nl                                 
J175854.2$-$290523&163&113 18285  390&18.00 &15.90 &   &   &8.55&7.90&0.6 &2.048&  \nl                                 
J175854.5$-$291031&164&113 18284  188&17.89 &15.62 &   &   &7.51&6.63&0.4 &1.922&  \nl                                 
J175854.5$-$290323&165&113 18286  189&17.51 &15.49 &   &   &8.85&8.67&0.2 &1.409&  \nl                                 
J175854.8$-$285831&167&113 18287 1897&14.43 &12.74 &   &   &6.84&6.94&0.25&1.403&  \nl                                 
J175854.9$-$290118&168&113 18286   24&9.71:& 9.40:&   &   &7.46&7.76&    &     & \nl                             
J175855.1$-$290037&169&113 18286   42&15.16 &13.59 &   &   &8.21&8.57&0.2 &1.718&  \nl                                 
J175855.1$-$290627&170&113 18285  564&17.79 &16.21 &   &   &6.99&5.53&0.15&2.127&  \nl                                 
J175855.4$-$285808&172&113 18287 1899&14.33 &13.13 &   &   &8.94&8.61&0.1 &1.240&  \nl                                 
J175855.7$-$291208&174&113 18283  170&17.40 &15.37 &   &   &7.86&7.67&0.25&1.615&  \nl                                 
J175855.7$-$290707&176&113 18285  138&17.08 &15.11 &   &   &8.00&7.73&0.2 &1.672&  \nl                                 
J175856.1$-$285943&183&113 18286   33&13.51:&12.24:&   &   &8.80&8.37&    &     &  \nl                            
J175856.4$-$290050&187&113 18286  454&19.08 &16.46 &   &   &6.73&5.37&0.3 &2.299:& \nl                              
J175856.7$-$290559&190&113 18285  133&16.93 &15.07 &   &   &8.68&8.37&0.15&1.448&  \nl                                 
J175856.9$-$290339&194&113 18286  166&17.34 &15.40 &   &   &8.32&8.20&0.2 &1.746&  \nl                                 
J175857.0$-$290445&197&113 18285  148&17.39 &15.17 &   &   &7.62&7.14&0.3 &1.649&  \nl                                 
J175857.1$-$290804&198&113 18284  189&16.93 &15.23 &   &   &8.91&8.80&0.1 &1.324&  \nl                                 
J175857.4$-$291215&200&113 18283  439&16.97 &15.88 &   &   &7.20&6.11&0.12&2.006&  \nl                                 
J175857.5$-$290537&202&113 18285   41&14.81 &13.43 &   &   &8.57&8.73&0.1 &1.993& \nl                                  
J175857.6$-$290632&204&113 18285   39&14.78 &13.41 &   &   &8.65&8.60&0.5 &1.776&  \nl                                 
J175857.8$-$290114&206&113 18286  193&17.73 &15.75 &   &   &6.65&5.34&0.4 &2.512:& \nl                               
J175858.0$-$290349&208&113 18285  420&18.28 &16.17 &   &   &8.46&7.84&0.6 &2.365:& \nl                               
J175858.4$-$285636&214&113 18287   41&15.38:&13.89:&   &   &8.40&7.35&    &     & \nl                              
J175858.5$-$291135&218&113 18284   54&14.91 &13.61 &   &   &9.01&8.81&0.1 &1.303&  \nl                                 
J175858.5$-$290845&219&113 18284  274&17.66 &15.71 &   &   &8.47&8.25&0.25&1.560&  \nl                                 
J175858.6$-$290722&220&113 18285   65&15.96 &14.23 &   &   &8.59&8.47&0.2 &1.691&  \nl                                 
J175900.0$-$285813&234&113 18287 1906&15.10 &13.74 &   &   &8.43&7.69&0.1 &1.229&  \nl                                 
J175900.2$-$290556&235&113 18285  156&16.68 &15.12 &   &   &9.27&8.97&0.1 &1.214&  \nl                                 
J175900.2$-$291111&236&113 18284   59&16.23 &14.50 &   &   &7.85&7.44&01.0&2.164&  \nl                                 
J175900.7$-$290305&242&113 18286  103&16.90 &14.95 &   &   &8.58&8.47&0.15&1.426&  \nl                                 
J175900.9$-$290952&244&113 18284  834&18.74 &16.67 &   &   &8.28&7.13&0.4 &1.912&  \nl                                 
J175900.9$-$291238&245&113 18283  201&17.26 &15.40 &   &   &8.35&8.19&0.2 &1.560&  \nl                                 
J175901.1$-$285821&247&113 18287 2391&17.48 &16.56 &   &   &5.96&4.46&    &     & Mira/nil; sat \nl                  
J175901.2$-$290910&249&113 18284  285&17.59 &15.61 &   &   &8.94&8.87&0.2 &1.396&  \nl                                 
J175901.2$-$290518&250&113 18285  296&17.89 &15.83 &   &   &8.07&7.29&0.15&1.800&  \nl                                 
J175901.4$-$290802&252&113 18284  491&18.34 &16.21 &   &   &7.98&7.03&0.3 &1.922&  \nl                                 
J175902.0$-$290531&258&113 18285  157&17.17 &15.22 &   &   &8.23&8.50&0.2 &1.555&  \nl                                 
J175902.2$-$290625&261&113 18285  134&17.36 &15.36 &   &   &7.55&6.60&0.3 &1.922&  \nl                                 
J175903.0$-$290137&266&113 18286  150&16.48 &15.07 &   &   &9.86&8.09&0.25&1.523& \nl                                  
J175903.0$-$290842&267&113 18284  474&17.17 &16.15 &   &   &8.39&8.32&0.2 &1.183&  \nl                                 
J175903.2$-$291205&270&113 18283   48&15.45 &13.94 &   &   &8.51&8.51&0.15&1.637&  \nl                                 
J175903.5$-$290603&273&113 18285  150&17.19 &15.30 &   &   &7.96&7.55&0.08&1.792&  \nl                                 
J175903.5$-$290830&274&113 18284   39&13.51:&12.49:&   &   &8.63&8.66&    &     & \nl                              
J175903.7$-$285919&277&113 18287 1968&16.56 &15.07 &   &   &8.77&8.79&0.15&1.444&  \nl                                 
J175903.8$-$291126&278&113 18284   36&13.38:&12.00: &   &   &8.27&8.41&    &     & \nl                            
J175904.0$-$290922&282&113 18284   37&13.57:&12.45:&   &   &8.38&8.43&    &     & \nl                              
J175904.2$-$291103&288&113 18284  192&17.59 &15.39 &   &   &8.32&8.05&0.5 &2.063&  \nl                                 
J175904.7$-$290745&291&113 18284  284&16.92 &15.43 &   &   &7.02&5.65&0.3 &2.103&  \nl                                 
J175905.2$-$290708&294&113 18285   47&14.95 &13.53 &   &   &8.64&8.81&0.1 &1.339&  \nl                                 
J175905.6$-$285833&299&113 18287 2109&18.47 &16.36 &   &   &7.95&6.86&0.3 &1.901&  \nl                                 
J175905.6$-$290545&302&113 18285   44&14.65 &13.37 &   &   &8.88&8.72&0.1 &1.273&  \nl                                 
J175905.6$-$290235&303&113 18286  399&18.48 &16.15 &   &   &7.08&6.01&0.3 &1.980& \nl                           
J175905.7$-$291128&304&113 18284  389&18.34 &16.38 &   &   &8.84&8.74&0.5 &1.303&  \nl                                 
J175905.8$-$291105&308&113 18284   66&15.45 &14.03 &   &   &8.13&8.09&0.1 &1.704&  \nl                                 
J175906.0$-$290620&311&113 18285   45&15.21 &13.71 &   &   &8.46&8.55&0.3 &1.776&  \nl                                 
J175906.0$-$290732&312&113 18285   60&14.32 &13.70 &   &   &8.73&8.63&0.06&1.409&  \nl                                 
J175906.5$-$290527&317&113 18285  107&16.59 &14.87 &   &   &8.04&8.04&0.15&1.459&  \nl                                 
J175907.2$-$291259&326&113 18283   37&14.66 &13.28 &   &   &8.25&7.93&0.4 &1.980&  \nl                                 
J175907.3$-$291241&327&113 18283  141&16.97 &15.12 &   &   &8.63&8.58&0.15&1.470& \nl                                  
J175907.3$-$291024&328&113 18284   57&16.01 &14.02 &   &   &7.24&6.68&0.6 &2.145&  \nl                                 
J175908.1$-$285827&336&113 18417 2005&16.99 &15.07 &   &   &8.43&8.22&0.25&1.486&  \nl                                 
J175908.3$-$290856&338&113 18414   98&17.19 &15.01 &   &   &7.46&6.40&0.6 &2.084&  \nl                                 
J175908.3$-$290729&339&113 18415   36&14.89 &13.42 &   &   &8.43&8.33&0.2 &1.555&  \nl                                 
J175908.6$-$291250&343&113 18413   45&15.73 &14.08 &   &   &8.25&8.11&0.1 &1.216&  \nl                                 
J175908.7$-$291345&345&113 18413 2532&19.05 &18.12 &   &   &7.86&7.73&0.35&1.968&  \nl                                 
J175909.3$-$290423&350&113 18415  116&16.72 &15.05 &   &   &8.93&8.85&0.15&1.365&  \nl                                 
J175909.5$-$290825&352&113 18414   49&16.09 &14.23 &   &   &7.48&7.31&0.2 &1.615&  \nl                                 
J175909.6$-$285800&353&113 18417 1984&15.90 &14.36 &   &   &8.98&8.83&0.1 &1.383&  \nl                                 
J175910.1$-$290936&357&113 18414   26&11.85:&11.00:&   &   &8.72&8.77&    &     &  \nl                             
J175910.1$-$291358&358&113 18413   59&16.34 &14.49 &   &   &8.32&7.72&0.35&1.746&     \nl                              
J175910.5$-$290129&363&113 18416  198&19.83 &16.44 &   &   &4.87&3.37&    &2.672:& Mira; few  \nl                     
J175910.6$-$290457&365&113 18415  117&17.54 &15.25 &   &   &7.24&6.45&0.4 &2.145&  \nl                                 
J175910.7$-$290035&370&113 18416   80&15.99 &14.24 &   &   &7.99&7.81&0.2 &1.724&  \nl                                 
J175910.9$-$290708&372&113 18415   43&15.37 &13.87 &   &   &8.15&8.07&0.25&1.637&  \nl                                 
J175910.9$-$285747&373&113 18417 1968&15.66 &14.00 &   &   &8.45&8.47&0.2 &     &  also long per \nl                         
J175911.1$-$290315&375&113 18416  597&18.27 &16.50 &   &   &6.71&5.34&0.5 &2.201:& \nl                               
J175911.1$-$291401&376&113 18413  410&18.35 &16.25 &   &   &7.87&6.95&0.35&2.110&  \nl                                 
J175911.3$-$290114&380&113 18416   29&10.04:& 9.64:&   &   &7.51&7.19&    &     & \nl                              
J175912.4$-$291357&388&113 18413   60&15.98 &14.27 &   &   &8.21&8.37&0.1 &1.448& \nl                                  
J175912.7$-$290608&392&113 18415  112&16.94 &15.02 &   &   &8.48&8.50&0.25&1.490&  \nl                                 
J175913.2$-$290901&395&113 18414  198&17.78 &15.62 &   &   &7.81&7.15&0.25&1.852&  \nl                                 
J175913.5$-$291409&399&113 18413  125&16.92 &15.00 &   &   &8.45&8.35&0.15&1.416&  \nl                                 
J175913.6$-$285812&400&113 18417 1956&14.18 &12.62 &   &   &7.36&7.10&0.4 &1.980&  \nl                                 
J175913.7$-$285852&402&113 18417 1965&19.49 &16.44 &   &   &6.96&5.75&4.0 &2.487:& Mira  \nl                         
J175913.8$-$290703&403&113 18415  100&16.78 &14.98 &   &   &8.78&8.32&0.15&1.514&  \nl                                 
J175914.0$-$290950&405&113 18414  551&18.27 &16.28 &   &   &7.14&5.58&0.3 &2.063&  \nl                                 
J175914.0$-$290815&406&113 18414  137&17.03 &15.12 &   &   &8.75&8.95&0.15&1.356&  \nl                                 
J175914.1$-$290241&407&113 18416  337&18.02 &15.98 &   &   &8.00&7.16&0.35&1.852&  \nl                                 
J175914.4$-$291335&410&113 18413   19&13.03:&11.80:&   &   &7.21&6.80&    &     & \nl                              
J175914.6$-$290853&412&113 18414   39&14.12 &13.54 &   &   &8.80&8.62&0.1 &1.433&  \nl                                 
J175914.8$-$291129&416&113 18414  338&18.13 &15.86 &   &   &7.40&6.41&0.4 &1.970&  \nl                                 
J175915.5$-$290134&421&113 18416 1364&18.05 &17.60 &   &   &6.80&6.07&0.4 &1.159& few \nl                              
J175915.5$-$290533&422&113 18415   74&17.16 &14.86 &   &   &7.60&6.80&0.5 &1.901&  \nl                                 
J175915.9$-$290824&425&113 18414  177&17.42 &15.33 &   &   &7.87&7.78&0.2 &1.816&  \nl                                 
J175915.9$-$290839&426&113 18414   61&16.16 &14.25 &   &   &7.64&7.59&0.25&1.724&  \nl                                 
J175916.0$-$290408&427&113 18415   71&16.36 &14.56 &   &   &8.66&8.53&0.15&1.738&  \nl                                 
J175916.4$-$290805&431&113 18414  211&17.35 &15.31 &   &   &7.57&6.75&0.25&1.808&  \nl                                 
J175916.8$-$290011&434&113 18416   28&9.75:& 9.42:&   &   &7.24&7.38&    &     & \nl                              
J175916.9$-$290225&435&113 18416  138&16.08 &14.83 &   &   &9.16&8.73&0.05&1.163& \nl                                  
J175917.0$-$290502&439&113 18415   39&14.43 &13.27 &   &   &9.04&8.85&0.2 &1.753&  \nl                                 
J175917.4$-$290643&442&113 18415   24&13.36:&12.04:&   &   &7.76&7.76&    &     &  \nl                             
J175917.4$-$291048&444&113 18414   56&16.55 &14.43 &   &   &7.58&6.93&0.7 &2.185:& \nl                               
J175917.9$-$290806&447&113 18414  290&18.22 &15.92 &   &   &7.53&6.48&0.3 &1.945&  \nl                                 
J175918.2$-$291434&452&113 18413   20&13.42:&11.98:&   &   &6.66&6.50&    &     & \nl                              
J175918.2$-$290123&453&113 18416   40&14.36 &12.64 &   &   &6.07&5.83&0.45&1.579&  \nl                                 
J175918.5$-$291302&454&113 18413   33&14.96 &13.70 &   &   &8.65&8.81&0.1 &2.145&  \nl                                 
J175918.5$-$290505&455&113 18415   73&16.73 &14.58 &   &   &7.29&6.58&0.8 &2.124:& \nl                               
J175918.5$-$290608&457&113 18415  415&18.73 &16.34 &   &   &7.32&5.77&0.4 &2.097:& \nl                               
J175918.6$-$290046&458&113 18416   49&14.75 &13.44 &   &   &8.75&8.74&0.15&1.399&  \nl                                 
J175918.9$-$290548&463&113 18415   62&15.76 &14.30 &   &   &8.94&8.64&0.08&1.209&  \nl                                 
J175919.6$-$290452&467&113 18415   26&14.17 &12.95 &   &   &8.50&8.42&0.15&1.776&  \nl                                 
J175919.8$-$291247&469&113 18413   56&15.65 &14.15 &   &   &8.77&8.80&0.15&1.448&  \nl                                 
J175920.1$-$291416&471&113 18413  670&17.93 &17.37 &   &   &8.48&8.42&1.0 &1.834&  \nl                                 
J175920.6$-$290953&477&113 18414 1489&19.00 &17.78 &   &   &9.06&8.10&0.9 &2.078&  \nl                                 
J175920.7$-$291506&478&113 18413  167&17.45 &15.34 &   &   &7.37&7.07&0.25&1.834&  \nl                                 
J175921.8$-$290841&486&113 18414   97&16.75 &14.84 &   &   &8.35&8.39&0.35&1.717&  \nl                                 
J175921.9$-$291058&487&113 18414   96&16.78 &14.86 &   &   &7.88&7.79&0.25&1.654&  \nl                                 
J175922.0$-$291229&488&113 18413   21&14.65 &12.84 &   &   &6.46&6.03&0.35&1.816&  \nl                                 
J175922.0$-$290547&490&113 18415   83&16.56 &14.69 &   &   &8.32&8.15&0.35&1.724& \nl                                  
J175922.1$-$291157&492&113 18413 1433&18.66 &17.49 &   &   &8.82&8.11&0.4 &1.944&  \nl                                 
J175922.5$-$291133&500&113 18414   95&16.57 &14.76 &   &   &8.80&8.68&0.25&1.792&  \nl                                 
J175923.2$-$290820&506&113 18414   34&14.71 &13.34 &   &   &8.54&8.31&0.3 &1.871&  \nl                                 
J175923.3$-$290902&510&113 18414   27&12.33:&10.78:&   &   &8.22&8.19&    &     & \nl                              
J175923.4$-$290216&511&113 18416  290&16.67 &15.56 &   &   &6.37&5.39&    &     & Mira; sat \nl                         
J175923.5$-$291451&512&113 18413  317&18.33 &16.08 &   &   &8.00&7.44&1.0 &2.117:& \nl                               
J175923.7$-$291236&514&113 18413 1776&18.94 &17.04 &   &   &6.83&5.64&~4  &2.405:& Mira  \nl                         
J175923.8$-$290952&516&113 18414   93&16.70 &14.94 &   &   &8.85&8.84&0.2 &1.490&  \nl                                 
J175924.4$-$291358&522&113 18413   31&15.54 &13.60 &   &   &7.24&6.21&0.4 &1.896&  \nl                                 
J175924.6$-$291237&524&113 18413  206&18.40 &15.88 &   &   &7.47&6.15&0.5 &2.020&  \nl                                 
J175925.3$-$290336&529&113 18416  521&18.40 &16.43 &   &   &7.13&5.70&0.25&2.110&  \nl                                 
J175925.7$-$290036&533&113 18416   58&15.20 &13.78 &   &   &8.96&8.77&0.15&1.594&  \nl                                 
J175926.1$-$290612&536&113 18545 1157&18.47 &17.06 &   &   &8.20&7.90&0.25&1.791&  \nl                                 
J175926.4$-$290707&540&113 18545   19&15.14 &13.34 &   &   &7.03&5.84&0.35&1.718&  \nl                                 
J175926.5$-$290218&541&113 18546  277&17.34 &15.46 &   &   &7.34&5.60&0.2 &1.891&\nl                                   
J175926.9$-$290345&545&113 18545  273&16.87 &15.63 &   &   &9.18&8.62&0.06&1.205&  \nl                                 
J175927.4$-$290310&550&113 18546  982&19.17 &17.80 &   &   &7.46&5.91&    &     & eclip bin \nl                        
J175927.7$-$291332&553&113 18543   48&16.14 &14.41 &   &   &8.39&8.23&0.1 &1.191&  \nl                                 
J175927.9$-$290530&556&113 18545  233&17.92 &15.71 &   &   &7.46&6.61&0.3 &1.861&  \nl                                 
J175927.9$-$291038&559&113 18544   65&16.85 &14.85 &   &   &7.76&7.62&0.3 &1.615& \nl                                  
J175928.0$-$290118&561&113 18546   36&14.49 &13.18 &   &   &8.38&8.19&0.1 &1.760&  \nl                                 
J175929.7$-$290926&575&113 18544   13&13.76 &12.64 &   &   &8.46&8.37&0.2 &2.164& sat? \nl                             
J175930.7$-$290950&581&113 18544   12&11.80:&11.60:&   &   &6.88&6.83&0.3 &     & \nl                              
J175931.1$-$290858&585&113 18544  399&18.04 &16.13 &   &   &7.05&5.67&0.3 &2.083:& \nl                              
J175932.1$-$290857&595&113 18544   26&15.51 &13.90 &   &   &8.32&7.67&0.3 &2.020&  \nl                                 
J175932.4$-$290802&598&113 18544  474&16.90 &16.07 &   &   &7.99&7.91&0.1 &1.311&  \nl                                 
J175934.0$-$290727&611&113 18545  927&17.38 &16.53 &   &   &8.64&8.57&0.06&1.110&  \nl                                 
J175935.0$-$291110&619&113 18544  255&16.54 &15.64 &   &   &8.26&7.64&0.08&1.154&  \nl                                 
J175935.2$-$290506&623&113 18545   81&14.79 &14.34 &   &   &8.56&8.74&0.05&1.980&  \nl                                 
J175936.1$-$290916&626&113 18544   21&15.60 &13.61 &   &   &6.96&6.75&0.4 &2.145&  \nl                                   
J175936.3$-$290516&628&113 18545  740&17.36 &16.38 &   &   &8.30&7.64&0.1 &1.079&  \nl                                 
J175937.0$-$290835&633&113 18544  131&17.46 &15.48 &   &   &7.44&6.56&0.6 &2.110&  \nl                                   
J175937.1$-$290253&634&113 18546  458&16.78 &16.04 &   &   &8.34&8.32&0.05&1.753&  \nl                                 
J175937.8$-$290925&639&113 18544   36&15.60 &14.18 &   &   &8.81&8.74&0.25&1.283&  \nl                                 
J175938.3$-$290129&644&113 18546  279&16.68 &15.66 &   &   &7.92&7.34&0.06&2.020&  \nl                                 
J175939.3$-$290915&655&113 18544   61&16.72 &14.69 &   &   &8.58&8.39&0.2 &1.649&  \nl                                 
J175940.9$-$290737&661&113 18545  532&17.03 &16.10 &   &   &8.41&8.36&0.06&1.968&  \nl                                 
J175941.9$-$290458&667&113 18545   33&13.95 &13.49 &   &   &8.25&8.28&0.03&1.039&  \nl                                 
J175943.4$-$290735&672&113 18545  228&16.25 &15.34 &   &   &7.88&7.03&0.05&1.451&  \nl                                 
J175947.0$-$290357&687&113 18675  756&17.17 &16.31 &   &   &7.65&7.02&0.25&1.079& \nl                                  
J175948.2$-$290350&689&113 18675 1135&17.62 &16.70 &   &   &7.05&6.37&0.15&2.033& definite mis-id \nl                  
J180232.1$-$300201&  6&119 19831   41&14.85 &13.71 &   &   &8.63&8.56&0.06&1.213& \nl                                  
J180233.9$-$300232&  8&119 19831  110&16.14 &14.99 &   &   &9.94&8.89&0.05&1.386&  \nl                                 
J180235.2$-$295855& 14&119 19832 2797&14.75 &12.73 &   &   &6.50&5.78&0.1 &1.951&  \nl                                 
J180236.1$-$300216& 17&119 19831  173&17.54 &15.62 &   &   &8.99&8.50&0.2 &1.422&  \nl                                 
J180236.6$-$295752& 19&119 19832 2843&16.95 &15.03 &   &   &7.81&7.54&0.2 &1.455& few  \nl                             
J180238.0$-$295933& 22&119 19832 2832&16.54 &14.78 &   &   &8.80&8.76&0.15&1.625&  \nl                                 
J180238.7$-$295954& 24&119 19831  471&18.18 &16.59 &   &   &7.75&6.25&0.35&2.075&  \nl                                 
J180239.4$-$295918& 25&119 19832 2808&15.22 &13.80 &   &   &8.93&8.78&0.1 &1.361&  \nl                                 
J180239.4$-$295636& 26&119 19832   18&11.96:&10.35:&   &   &7.46&7.16&    &     & \nl                              
J180240.2$-$295821& 27&119 19832 2805&15.40 &13.87 &   &   &7.96&7.78&0.15&1.625&  \nl                                 
J180240.6$-$300053& 28&119 19831   75&16.51 &14.67 &   &   &8.16&8.02&0.15&1.433&  \nl                                 
J180241.0$-$295902& 31&119 19832 2872&17.83 &15.80 &   &   &8.80&8.43&0.2 &2.060&  \nl                                 
J180241.7$-$295753& 35&119 19832 2804&15.38 &13.83 &   &   &8.54&8.27&0.2 &1.716& few \nl                              
J180241.8$-$295957& 36&119 19831  139&17.40 &15.52 &   &   &8.25&7.91&0.3 &1.990&  \nl                                 
J180242.1$-$295937& 38&119 19832 2794&12.25:&11.25:&   &   &8.28&8.19&    &     & \nl                              
J180242.9$-$300335& 41&119 19831  494&19.06 &16.88 &   &   &7.48&5.86&0.5 &2.106&\nl                                   
J180243.5$-$295615& 45&119 19832   45&16.88 &15.02 &   &   &8.36&8.25&0.25&1.625&  \nl                                 
J180245.1$-$295813& 53&119 19832 2824&16.01 &14.52 &   &   &9.28&8.99&0.1 &1.256&  \nl                                 
J180245.4$-$295536& 57&119 19833   45&14.57 &13.51 &   &   &8.40&8.25&0.5 &1.249& \nl                                  
J180245.6$-$300328& 60&119 19831   49&15.62 &14.14 &   &   &8.99&8.86&0.06&1.703&  \nl                                 
J180245.8$-$300111& 61&119 19831  299&18.22 &16.22 &   &   &7.58&6.00&0.35&2.060& BMB3 7 \nl                           
J180248.4$-$300309& 70&119 19961  106&17.46 &15.40 &   &   &8.12&7.69&0.3 &1.759&\nl                                   
J180249.0$-$295430& 76&119 19963  151&17.75 &15.65 &   &   &8.00&7.02&0.4 &1.899&\nl                                   
J180249.5$-$295852& 79&119 19962 2779&16.22 &14.63 &   &   &8.53&8.06&0.3 &1.798& BMB6 6 \nl                           
J180251.1$-$300326& 83&119 19961   18&12.74:&12.47:&   &   &8.31&8.25&    &     & \nl                              
J180251.2$-$300013& 85&119 19961   64&17.30 &15.14 &   &   &7.49&6.61&0.7 &1.943& BMB7 7 \nl                           
J180251.8$-$300246& 88&119 19961  132&17.41 &15.32 &   &   &8.11&7.65&0.25&1.767&  \nl                                 
J180252.4$-$300023& 90&119 19961  163&17.80 &15.70 &   &   &8.57&7.87&0.3 &2.060& BMB12 6 \nl                          
J180252.7$-$295459& 92&119 19963  178&17.78 &15.76 & 9.64& 8.47&8.72&8.65&0.2 &1.555& BMB11 7; few  \nl                     
J180252.9$-$300106& 94&119 19961  146&18.05 &15.92 &   &   &8.36&7.63&0.4 &2.004& BMB10 6.5 \nl                        
J180253.0$-$300249& 96&119 19961  162&17.60 &15.53 &   &   &7.51&6.66&0.3 &1.920&  \nl                                 
J180253.8$-$295425&102&119 19963   60&16.10 &14.48 &   &   &8.25&8.16&0.15&1.554& BMB15 6 \nl                          
J180254.1$-$300048&103&119 19961   84&17.03 &14.89 &   &   &7.65&6.65&0.5 &1.899& BMB12 6 \nl                          
J180256.1$-$295534&110&119 19963  138&18.08 &15.73 &   &   &6.71&5.77&0.5 &2.447:& \nl                               
J180256.7$-$295705&114&119 19962   23&15.11 &13.63 &   &   &8.47&8.14&0.5 &1.869&  \nl                                 
J180256.9$-$295548&118&119 19962   99&16.79 &15.71 &   &   &7.63&7.32&0.15&1.444& BMB20/21 6/6.5\nl                    
J180257.2$-$295201&121&119 19963   67&16.07 &14.54 &   &   &9.24&8.75&0.1 &1.631&  \nl                                 
J180257.4$-$300351&126&119 19960   22&13.97 &12.77 &   &   &8.27&7.69&0.1 &1.625&  \nl                  
J180257.6$-$295124&128&119 19964   19&13.07:&12.25:&   &   &6.80&6.32&    &     &  \nl                            
J180258.1$-$295049&131&119 19964   25&14.47 &13.17 &   &   &8.36&8.14&0.15&1.365&  \nl                                 
J180258.4$-$300311&135&119 19961  108&17.20 &15.26 &   &   &8.38&8.49&0.2 &1.767&  \nl                      
J180258.8$-$295426&136&119 19963   30&15.29 &13.77 &   &   &8.23&6.95&0.5 &2.581:& dp(0.2,1.869)   \nl     
J180258.9$-$295221&138&119 19963   32&14.86 &13.57 &   &   &8.88&8.77&0.08&1.554&  \nl                                 
J180258.9$-$300108&139&119 19961   71&16.64 &14.87 &   &   &8.70&8.56&0.25&1.554& BMB27 6.5 \nl                        
J180259.0$-$295757&140&119 19962 2777&16.28 &14.45 &   &   &8.01&7.40&0.3 &2.004&  \nl                                 
J180259.6$-$300253&147&119 19961  176&17.71 &15.64 & 8.47& 7.23&7.03&5.28&0.4 &2.371:& BMB28 7 \nl                     
J180300.2$-$295514&152&119 19963   15&12.27:&10.84:&   &   &7.81&7.55&    &     & \nl                              
J180300.6$-$295018&156&119 19964   22&14.40 &12.93 &   &   &8.27&7.63&0.05&1.919&  \nl                                 
J180301.1$-$300142&162&119 19961   20&15.33 &13.07 &   &   &6.36&5.51&0.1 &2.004&  \nl                                 
J180301.7$-$295959&166&119 19961  155&17.58 &15.63 &   &   &8.31&8.09&0.25&1.527& BMB35 6.5 \nl                        
J180301.7$-$295053&167&119 19964   38&14.62 &13.59 &   &   &9.02&8.60&0.1 &1.103&  \nl                                 
J180302.6$-$295645&172&119 19962   28&16.30 &14.43 &   &   &7.96&7.44&0.6 &2.091& BMB37 6 \nl                          
J180303.2$-$295515&179&119 19963   20&14.24 &12.83 &   &   &8.48&8.60&0.06&1.823&  \nl                                 
J180303.8$-$300242&183&119 19961  130&17.25 &15.31 &   &   &8.50&7.88&0.5 &1.850& BMB36 6.5 \nl                        
J180304.0$-$295135&184&119 19964   79&16.35 &14.71 &   &   &9.03&8.55&0.1 &2.090& BMB44 6.5 \nl                        
J180304.6$-$300405&187&119 19960   39&15.53 &14.15 &   &   &8.60&8.32&0.2 &1.645&  \nl                                 
J180304.9$-$295258&189&119 20093 2040&17.18 &15.29 &   &   &8.43&8.31&0.15&1.573&  \nl                                 
J180305.3$-$295516&192&119 20093 2054&17.71 &15.51 & 8.35& 7.00&6.86&5.52&0.3 &2.060& BMB46 7  \nl                         
J180305.4$-$295033&194&119 20094 1986&13.75 &12.58 &   &   &8.06&7.68&0.05&1.204&   \nl                                
J180305.9$-$295345&197&119 20093 2027&16.37 &14.78 &   &   &9.09&8.48&0.4 &1.824&\nl                                   
J180305.9$-$300508&198&119 20090   51&16.70 &14.91 &   &   &8.87&8.83&0.2 &1.370&  \nl                                 
J180306.3$-$295204&205&119 20093 2179&18.38 &16.42 &   &   &7.40&5.75&0.2 &2.142&\nl                                   
J180306.3$-$295141&206&119 20094 2053&17.29 &15.29 &   &   &7.82&6.75&0.4 &1.910& BMB50 7\nl                           
J180306.7$-$300136&207&119 20091    8&12.47:&11.02:&   &   &8.62&8.62&   &     & \nl                              
J180307.0$-$300633&211&119 20090  106&18.04 &16.00 &   &   &7.93&7.09&0.2 &1.888& BMB45 7 \nl                          
J180307.1$-$300519&213&119 20090  118&18.26 &16.08 &   &   &8.60&7.55&0.4 &1.866& BMB49 6.5 \nl                        
J180307.4$-$300255&215&119 20091   11&14.60 &13.29 &   &   &8.45&8.31&0.25&1.609&   \nl                                
J180307.8$-$300452&217&119 20090   26&15.15 &13.88 &   &   &9.16&8.82&0.05&2.075&  \nl                                 
J180308.2$-$300330&221&119 20091   95&17.23 &15.64 &   &   &8.22&7.71&0.2 &1.966& BMB53 6.5 \nl                        
J180308.5$-$300525&224&119 20090  355&18.97 &16.68 & 8.41& 6.94&6.32&4.82&4.0 &2.531:& Mira; BMB54 7 \nl                  
J180308.8$-$295220&226&119 20093 2095&18.17 &16.02 &   &   &7.83&6.43&0.3 &1.991& BMB59 6.5 \nl                        
J180308.9$-$300551&227&119 20090   43&16.06 &14.64 &   &   &9.45&8.75&0.1 &1.292&   \nl                                
J180309.3$-$295241&231&119 20093 1994&15.15 &13.71 &   &   &8.04&7.97&0.2 &1.631&  \nl                                 
J180309.4$-$300241&234&119 20091   13&14.55 &13.45 &   &   &8.86&8.83&0.1 &1.292&  \nl                                 
J180310.0$-$300138&236&119 20091   27&15.95 &14.51 &   &   &8.66&8.45&0.1 &1.473& BMB58 6 \nl                          
J180310.6$-$295619&240&119 20092  756&15.85 &14.66 &   &   &9.30&8.49&0.15&1.079&  \nl                                 
J180311.2$-$295312&242&119 20093 1991&14.41 &13.01 &   &   &8.09&7.79&0.6 &1.850&\nl                                   
J180311.9$-$295900&248&119 20092 2338&13.89 &12.74 &   &   &9.09&8.73&0.05&1.598& \nl                                  
J180312.5$-$300429&254&119 20090   55&17.48 &15.23 &   &   &7.04&5.98&0.4 &1.954& BMB54 7\nl                           
J180313.4$-$300056&256&119 20091  153&19.23 &16.41 &   &   &7.38&6.35&0.7 &2.346:& BMB67 8 \nl                     
J180313.9$-$295620&259&119 20092  748&15.93 &14.16 &   &   &8.02&7.36&0.4 &1.876&  \nl                                 
J180317.9$-$300230&292&119 20091 3890&17.02 &15.29 &   &   &8.86&8.58&0.3 &2.060& BMB79 6.5 \nl                        
J180318.1$-$300309&294&119 20091 3853&16.03 &14.51 &10.10& 9.09&9.30&8.85&0.1 &1.943& BMB84 6 \nl                          
J180318.4$-$295346&299&119 20093   55&17.57 &15.81 & 8.64& 7.29&7.15&5.29&0.2 &1.879& BMB86 9; long per?  \nl               
J180320.1$-$295935&313&119 20092 4026&15.32 &13.83 &   &   &8.64&8.35&0.1 &1.088& BMB89 5 \nl                          
J180320.3$-$295432&317&119 20093   31&16.36 &14.79 & 9.41& 8.37&8.62&8.57&0.1 &1.389& BMB91 6.5 \nl                        
J180320.8$-$300451&319&119 20090 3751&17.29 &15.42 &   &   &9.12&8.66&0.15&1.396&  \nl                                 
J180322.3$-$300255&333&119 20091 3839&14.88 &13.22 & 8.23& 7.18&7.43&7.08&0.25&1.710& BMB93 6 \nl                          
J180323.4$-$300838&339&119 20219   52&16.97 &14.97 &   &   &7.58&6.20&1.0 &2.091&  \nl                                 
J180323.9$-$295410&346&119 20223  112&17.18 &15.15 &   &   &8.08&6.92&0.35&1.943& BMB103 6.5\nl                        
J180323.9$-$300004&347&119 20221   80&15.73 &14.33 &10.04& 9.04&9.14&8.74&0.5 &2.602:& B28 5; dp(0.1,1.7)\nl 
J180324.0$-$295925&350&119 20222 2570&17.59 &15.57 &   &   &8.16&7.97&0.3 &1.790& BMB101 6.5\nl                        
J180324.5$-$300414&354&119 20220   61&16.18 &14.41 & 9.32& 8.27&8.71&8.83&0.1 &1.518& BMB99 6.5 \nl                        
J180325.1$-$300849&357&119 20219   54&15.13 &14.14 &   &   &9.22&8.56&0.03&1.370&  \nl                                 
J180325.3$-$300645&360&119 20220   54&16.22 &14.37 &   &   &8.43&8.56&0.4 &1.790& BMB102 6.5   \nl                     
J180325.3$-$295947&361&119 20221  104&16.79 &14.70 & 9.36& 8.29&8.51&8.46&0.15&1.573& BMB106 6  \nl                        
J180325.8$-$295847&365&119 20222 2546&16.83 &14.84 & 8.54& 7.51&7.54&6.89&0.7 &2.060& BMB108 6 \nl                         
J180326.3$-$295653&368&119 20222   12&13.44:&12.52:&   &   &8.27&8.25&   &     & \nl                              
J180326.4$-$300700&370&119 20220  395&16.98 &16.02 &   &   &8.91&8.38&0.15&1.455&  \nl                                 
J180327.3$-$300102&375&119 20221  126&16.91 &14.81 & 8.80& 7.65&7.78&7.10&0.3 &1.824& BMB114 6.5\nl                        
J180327.5$-$300224&379&119 20221   55&15.23 &13.65 & 9.09& 8.09&8.29&8.71&0.1 &1.220& B47 5 \nl                            
J180327.7$-$300655&381&119 20220  158&16.75 &15.41 &   &   &9.50&8.37&0.15&2.107&  \nl                                 
J180328.4$-$295545&382&119 20222   81&18.30 &16.02 &   &   &7.25&5.54&0.1 &2.030& BMB119 9 \nl                         
J180328.9$-$300334&386&119 20221   30&12.74:&11.28:&   &   &8.81&8.90&   &     & \nl                              
J180329.3$-$300248&387&119 20221   45&14.16 &12.97 &   &   &8.61&8.72&0.35&1.767& B54 1 bl? \nl                        
J180329.4$-$295939&389&119 20222 2540&16.38 &15.11 & 8.58& 7.31&7.38&5.99&0.1 &2.075& BMB120 7 \nl                         
J180329.6$-$300108&394&119 20221  108&16.51 &14.63 &   &   &7.92&7.25&0.5 &1.943& BMB121 6.5   \nl                     
J180330.0$-$295821&397&119 20222 2501&14.71 &13.30 &   &   &8.43&8.21&0.25&1.603& B61 5 \nl                            
J180331.0$-$295846&405&119 20222 2498&13.16:&12.08:&   &   &9.40&8.49&   &     & \nl                              
J180331.1$-$295908&407&119 20222 2573&17.61 &15.46 & 9.26& 8.12&8.41&7.51&0.2 &1.806& BMB127 7  \nl                        
J180331.2$-$295333&409&119 20223  158&17.29 &15.35 & 8.77& 7.56&7.75&7.64&0.2 &1.910& BMB136 8 \nl                         
J180331.3$-$300101&410&119 20221   39&15.02 &13.10 &   &   &6.71&6.33&0.2 &1.625& BMB129 6.5 \nl                       
J180331.6$-$300043&413&119 20221   44&14.20 &13.03 & 9.58& 8.74&8.98&8.77&0.06&1.568& B66 3   \nl                          
J180331.9$-$300608&414&119 20220   50&16.35 &14.39 & 8.93& 7.85&8.01&7.36&0.3 &2.060& BMB128 7  \nl                        
J180331.9$-$300027&415&119 20221   56&15.35 &13.65 & 8.81& 7.79&8.17&8.13&0.25&1.638& BMB131 6 \nl                         
J180332.3$-$300147&418&119 20221   60&15.20 &13.71 &   &   &8.91&8.60&0.1 &2.031& BMB133 6 \nl                         
J180332.3$-$300444&419&119 20220   81&16.34 &14.61 & 9.77& 8.71&8.82&8.73&0.3 &1.416& BMB133 6   \nl                       
J180333.2$-$295910&425&119 20222 2518&17.19 &15.12 &   &   &8.37&7.56&1.0 &2.316:& BMB140 6.5 \nl                   
J180333.4$-$300523&426&119 20220   89&15.98 &14.73 & 9.23& 8.06&8.86&7.39&0.2 &1.869& BMB134 7  \nl                        
J180334.1$-$295957&432&119 20221  178&17.52 &15.30 & 8.17& 6.90&6.97&5.56&0.35&2.075& BMB142 8  \nl                        
J180334.2$-$300104&433&119 20221   88&16.34 &14.39 &   &   &7.71&7.15&0.2 &2.004& BMB143 6.5 \nl                       
J180334.6$-$300137&437&119 20221  100&16.48 &14.57 &   &   &8.36&8.35&0.2 &1.760& BMB146 6; few   \nl                   
J180336.9$-$300148&445&119 20221  290&16.23 &15.42 & 8.51& 7.21&7.07&6.03&0.2 &1.043& BMB152 9  \nl                        
J180339.1$-$295826&456&119 20222 2502&14.77 &13.55 &   &   &9.23&8.99&0.1 &1.737& B101 4 \nl                           
J180340.2$-$295531&459&119 20223   43&14.75 &13.35 &   &   &8.04&8.24&0.2 &1.473&  \nl                                 
J180340.4$-$295612&460&119 20222   19&15.80 &14.40 &   &   &9.26&8.91&0.15&1.258&  \nl                                 
J180342.9$-$295606&467&119 20352   15&10.41:&9.94:&   &   &8.62&8.62&    &     & BMB156 6 \nl                     
J180345.1$-$295516&473&119 20353  289&16.73 &15.77 &   &   &0.44&8.91&0.05&1.412&  \nl                                 
J180346.2$-$295912&476&119 20352 2239&17.60 &15.41 & 8.20& 6.87&7.00&5.21&0.3 &2.075& BMB179 7 \nl                         
J180348.5$-$295946&483&119 20351   63&16.77 &14.60 &   &   &7.79&7.09&0.25&1.850& BMB186 6.5\nl                        
J180350.9$-$295618&491&119 20352   38&18.76 &16.79 & 7.61& 6.44&6.31&5.26&4.0 &2.498:& BMB194 6.5; Mira \nl               
\enddata
\tablenotetext{A}{ The full ISOGAL designation is of the form:
ISOGALP~Jhhmmss.s-ddmmss (2000). P signifies `provisional'.}

\tablenotetext{B}{ BW denotes the running number during the analyses 
of the Sgr I and NGC6522 fields.}

\tablenotetext{C}{ The MACHO 3-digit identifier refers to the field, tile,
and the sequence number (see Alcock et al.~1999).}

\tablenotetext{D}{$V$ and $R$ mags are flux-weighted time-averages. Those marked
with ``:'' are by-eye estimates; these MACHO lightcurves show saturation effects.
In these cases, no estimates for
Amp.\ and $\log P$ are given.}

\tablenotetext{E}{Periods are in days.  Those marked with ``:'' are by-eye
estimates for some long-period variables as described in the text.} 

\tablenotetext{F}{BMB denotes Blanco McCarthy Blanco (1984),
B denotes Blanco (1986), and the following number is the
M giant spectral sub$-$type. Those marked with ``Mira'' are also
listed in Table~1.  In some cases we note a second or ``double'' period 
using the notation: dp(Amp.\ ,$\log P$).} 

\end{deluxetable}            
\end{landscape}

\clearpage
\textheight22cm
\textwidth17cm
\hoffset-0.6in
\tighten
 
\begin{deluxetable}{llllll}
\tablewidth12cm
\tablecaption{Mira variables in the survey area}
\tablehead{
\colhead{Name} &
\colhead{MACHO} &
\colhead{[7]$_{avg}$} &
\colhead{[15]} &
\colhead{$K_{avg}$} &
\colhead{Period$^1$} 
}
\startdata
\multicolumn{6}{c}{NGC\,6522}                                       \nl
TLE D9 & missing$^2$     & 4.96        & 3.64 &  5.78    &          \nl       
TLE228 & missing$^2$     & 6.03        & 5.11 &  6.60    &          \nl      
TLE403 & 119 20090 355   & 6.32        & 4.82 &  7.00    &340/335   \nl      
TLE238 & missing$^2$     & 6.84        & 5.80 &  7.44    &          \nl       
TLE136 & 119 20352 38    & 6.31        & 5.26 &  6.44    &315/270   \nl       
\multicolumn{6}{c}{SGR\,I}                                          \nl       
TLE65  & 113 18155 43    & 6.33        & 5.28 & 6.98     &296/265   \nl       
TLE79  & 113 18287 2391  & 5.96        & 4.46 & 6.66     &mis-id$^4$\nl       
TLE53  & 113 18416 198   & 4.87        & 3.37 & 6.43     &470/500::$^3$\nl    
TLE87  & 113 18417 1965  & 6.96        & 5.75 & 7.29     &307/315   \nl       
TLE54  & 113 18416 290   & 6.37        & 5.39 & 7.04     &mis-id$^4$\nl       
TLE39  & 113 18413 1776  & 6.83        & 5.64 & 7.80     &254/235   \nl       
TLE55  & missing$^2$     & 6.51        & 5.47 & 6.81     & -        \nl       
TLE57  & missing$^2$     & 7.24        & 6.97 & 8.18     & -        \nl       
TLE56  & 113 18675 1135  & 7.05        & 6.37 & 7.43     &mis-id$^4$\nl       
                                                                              
\enddata                                                                 
                                                                              
\tablenotetext{}{Notes:}                                                                        
\tablenotetext{1}{The left period was derived from MACHO data and the right
by Lloyd Evans (1976).}                
                                                              
\tablenotetext{2}{``missing" stars probably appeared constant or non-stellar 
in the MACHO  data due to saturation, and were rejected before the matching 
stage.}          
                                                                              
\tablenotetext{3}{ fragmentary MACHO light curve.}                                           
                                                                              
\tablenotetext{4}{ An apparent MACHO counterpart was mistakenly found 
during the cross-identification process; the image of the correct star was 
probably saturated.}

\end{deluxetable}            

\clearpage
\pagestyle{empty}
\textheight22cm
\textwidth17cm
\hoffset-0.6in
\tighten
 
\begin{landscape}
\begin{deluxetable}{llllllllrl}
\tablewidth18cm
\tablecaption{Spectral Energy Distribution Modeling Results}
\tablehead{
\colhead{BW} &
\colhead{MACHO} &
\colhead{FW87} &
\colhead{M$_{\rm bol}$} &
\colhead{T$_{(V-K)}$ } &
\colhead{T$_{mod}$} &
\colhead{\.{M}$_{\rm L4}$} &
\colhead{\.{M}} &
\colhead{x[15]} &
\colhead{V$_{exp}$} 
}
\startdata
 92 &119 19963  178& 11     & $-$3.12 & 3077 &	2500 &	1.0e$-$7 & 2.3e$-$8 &    0.79 & 13.6 \nl
147 &119 19961  176& 28     & $-$4.35 & 2912 &	3000 &	8.2e$-$7 & 4.4e$-$7 &  129.65 & 16.2 \nl
192 &119 20093 2054& 46	    & $-$4.47 & 2882 &	3000 &	8.2e$-$7 & 4.8e$-$7 &   92.79 & 16.6 \nl
294 &119 20091 3853& 84	    & $-$2.69 & 3442 &	3500 &	6.8e$-$8 & 1.2e$-$8 &    2.23 & 14.4 \nl
299 &119 20093   55& 86	    & $-$4.18 & 2939 &	3000 &	8.2e$-$7 & 3.9e$-$7 &  131.36 & 15.6 \nl
317 &119 20093   31& 91	    & $-$3.37 & 3273 &	3200 &	5.4e$-$8 & 1.5e$-$8 &    0.72 & 15.1 \nl
333 &119 20091 3839& 93	    & $-$4.53 & 3318 &	3350 &	6.3e$-$8 & 3.8e$-$8 &    9.51 & 21.3 \nl
347 &119 20221   80& B28    & $-$2.84 & 3489 &	3500 &	1.4e$-$7 & 2.6e$-$8 &    2.27 & 14.8 \nl
354 &119 20220   61& 94	    & $-$3.39 & 3286 &	3350 &	6.3e$-$8 & 1.7e$-$8 &   -0.37 & 16.4 \nl
361 &119 20221  104& 106    & $-$3.38 & 3197 &	3200 &	1.1e$-$7 & 3.0e$-$8 &    0.80 & 15.1 \nl
365 &119 20222 2546& 108    & $-$4.19 & 3076 &	2500 &	1.9e$-$7 & 9.1e$-$8 &   17.36 & 16.3 \nl
375 &119 20221  126& 114    & $-$3.98 & 3085 &	2500 &	1.9e$-$7 & 7.9e$-$8 &   14.74 & 15.5 \nl
379 &119 20221   55& B47    & $-$3.69 & 3408 &	3350 &	3.1e$-$8 & 1.1e$-$8 &   -2.70 & 17.6 \nl
389 &119 20222 2540& 120    & $-$4.20 & 3113 &	3000 &	4.0e$-$7 & 1.9e$-$7 &   61.22 & 16.1 \nl
407 &119 20222 2573& 127    & $-$3.43 & 3051 &	2500 &	1.9e$-$7 & 5.4e$-$8 &   12.01 & 13.7 \nl
409 &119 20223  158& 136    & $-$3.98 & 3016 &	2500 &	1.9e$-$7 & 7.9e$-$8 &    2.58 & 15.5 \nl
413 &119 20221   44& B66    & $-$3.38 & 3805 &	3750 &	4.2e$-$8 & 1.2e$-$8 &    1.40 & 18.7 \nl
415 &119 20221   56& 131    & $-$3.95 & 3339 &	3350 &	6.3e$-$8 & 2.6e$-$8 &    0.98 & 18.6 \nl
414 &119 20220   50& 128    & $-$3.82 & 3197 &	3200 &	2.1e$-$7 & 7.8e$-$8 &   11.26 & 15.6 \nl
419 &119 20220   81& 133    & $-$2.95 & 3329 &	3350 &	1.3e$-$8 & 2.7e$-$8 &    0.84 & 14.7 \nl
426 &119 20220   89& 134    & $-$3.58 & 3284 &	3350 &	2.5e$-$7 & 7.9e$-$8 &   17.29 & 15.9 \nl
432 &119 20221  178& 142    & $-$4.58 & 2894 &	3000 &	8.2e$-$7 & 5.2e$-$7 &   91.57 & 17.1 \nl
445 &119 20221  290& 152    & $-$4.27 & 3120 &	3000 &	4.0e$-$7 & 2.0e$-$7 &   50.96 & 16.4 \nl
476 &119 20352 2239& 179    & $-$4.59 & 2880 &	3000 &	8.2e$-$7 & 5.2e$-$7 &  139.46 & 17.1 \nl
491 &119 20352   38& 194    & $-$5.06 & 2718 &	3000 &	3.3e$-$6 & 2.9e$-$6 &  104.12 & 16.3 \nl
224 &119 20090  355& TLE403 & $-$4.51 & 2741 &	3000 &	3.3e$-$6 & 2.0e$-$6 &  186.07 & 14.3 \nl
\enddata
\end{deluxetable}            
\end{landscape}

\end{document}